\documentclass[iop]{emulateapj}

\usepackage{natbib}
\usepackage{color}
\usepackage{ulem}

\usepackage[T1]{fontenc}

\usepackage{bm}

\newcommand{\eg}{e.g., }
\newcommand{\ie}{i.e., }
\newcommand{\Msun}{M_{\odot}}
\newcommand{\Rsun}{R_{\odot}}
\newcommand{\kms}{km~s$^{-1}$}

\newcommand{\Mej}{M_{\rm ej}}

\newcommand{\dmdt}{$|\Delta m/ \Delta t|$}
\def\gsim{\mathrel{\rlap{\lower 4pt \hbox{\hskip 1pt $\sim$}}\raise 1pt
\hbox {$>$}}}
\def\lsim{\mathrel{\rlap{\lower 4pt \hbox{\hskip 1pt $\sim$}}\raise 1pt
\hbox {$<$}}}

\def\ion#1#2{{\rm #1}~{\sc #2}}

\shorttitle{Rapidly Rising Transients from Subaru/HSC Transient Survey}
\shortauthors{Tanaka, M., et al.}

\begin{document}

\title{Rapidly Rising Transients from Subaru Hyper Suprime-Cam Transient Survey\footnote{Based [in part] on data collected at Subaru Telescope, which is operated by the National Astronomical Observatory of Japan}}
\author{
  Masaomi Tanaka\altaffilmark{1,2},
  Nozomu Tominaga\altaffilmark{3,2},
  Tomoki Morokuma\altaffilmark{4,2},
  Naoki Yasuda\altaffilmark{2},
  Hisanori Furusawa\altaffilmark{1},\\
  Petr V. Baklanov\altaffilmark{5,6,7},
  Sergei I. Blinnikov\altaffilmark{5,2,8},
  Takashi J. Moriya\altaffilmark{9},
  Mamoru Doi\altaffilmark{4,2},
  Ji-an Jiang\altaffilmark{4},\\
  Takahiro Kato\altaffilmark{10},
  Yuki Kikuchi\altaffilmark{4},
  Hanindyo Kuncarayakti\altaffilmark{11,12},
  Tohru Nagao\altaffilmark{13},\\
  Ken'ichi Nomoto\altaffilmark{2,14},
  and Yuki Taniguchi\altaffilmark{4}
}

\altaffiltext{1}{National Astronomical Observatory of Japan, Mitaka, Tokyo 181-8588, Japan; masaomi.tanaka@nao.ac.jp}
\altaffiltext{2}{Kavli Institute for the Physics and Mathematics of the Universe (WPI), The University of Tokyo, Kashiwa, Chiba 277-8583, Japan}
\altaffiltext{3}{Department of Physics, Faculty of Science and Engineering, Konan University, Kobe, Hyogo 658-8501, Japan}
\altaffiltext{4}{Institute of Astronomy, Graduate School of Science, The University of Tokyo, Mitaka, Tokyo 181-0015, Japan}
\altaffiltext{5}{Institute for Theoretical and Experimental Physics (ITEP), Bolshaya Cheremushkinskaya 25, 117218 Moscow, Russia}
\altaffiltext{6}{Novosibirsk State University, Novosibirsk 630090, Russia}
\altaffiltext{7}{National Research Nuclear University MEPhI, 115409 Moscow, Russia}
\altaffiltext{8}{All-Russia Research Institute of Automatics (VNIIA), 127055 Moscow, Russia}
\altaffiltext{9}{Argelander Institute for Astronomy, University of Bonn Auf dem H{\"u}gel 71, D-53121 Bonn, Germany}

\altaffiltext{10}{Department of Physics, The University of Tokyo, Tokyo 113-0033, Japan}
\altaffiltext{11}{Millennium Institute of Astrophysics, Casilla 36-D, Santiago, Chile}
\altaffiltext{12}{Departamento de Astronom\'ia, Universidad de Chile, Casilla 36-D, Santiago, Chile}
\altaffiltext{13}{Research Center for Space and Cosmic Evolution, Ehime University, Matsuyama 790-8577, Japan}
\altaffiltext{14}{Hamamatsu Professor}

\begin{abstract}
  We present rapidly rising transients discovered by a high-cadence
  transient survey with Subaru telescope and Hyper Suprime-Cam.
  We discovered five transients at $z=0.384-0.821$
  showing the rising rate faster than 1 mag per 1 day in the restframe
  near-ultraviolet wavelengths.
  The fast rising rate and brightness are the most similar to SN 2010aq and PS1-13arp,
  for which the ultraviolet emission within a few days
  after the shock breakout was detected.
  The lower limit of the event rate of rapidly rising transients is
  $\sim 9 \%$ of core-collapse supernova rates,
  assuming a duration of rapid rise to be 1 day.
  We show that the light curves of the three faint objects agree with
  the cooling envelope emission from the explosion of red supergiants.
  The other two luminous objects are, however, 
  brighter and faster than the cooling envelope emission.
  We interpret these two objects 
  to be the shock breakout from dense wind with the mass loss rate of
  $\sim 10^{-3}\ \Msun$ yr$^{-1}$, as also proposed for PS1-13arp.
  This mass loss rate is higher than that typically observed for
  red supergiants.
  The event rate of these luminous objects is $\gsim 1 \%$ of
  core-collapse supernova rate, and thus, our study implies that
  more than $\sim 1 \%$ of massive stars can experience an
  intensive mass loss at a few years before the explosion.
\end{abstract}


\section{Introduction}
\label{sec:introduction}

The transient sky has been intensively explored
by various surveys in the last decade.
Especially, optical surveys using wide-field cameras,
such as Palomar Transient Factory \citep[PTF,][]{law09,rau09},
Catalina Real-Time Transient Survey \citep[CRTS,][]{drake09},
and Pan-STARRS1 \citep[PS1, \eg][]{kaiser10},
have significantly contributed to building our knowledge
on the transient phenomena in the Universe.

One of the important discovery spaces for transient surveys
is phenomena with a short timescale, \ie $\lsim 1$ day.
There are, in fact, several theoretical expectations for
such short-timescale transients.
For supernovae (SNe), shock breakout emission
should have timescale of $\sim 1$ hr
for the case of red supergiant progenitors
\citep[\eg][]{falk78,klein78sbo,matzner99}.
The subsequent cooling emission lasts for a few days
\citep[\eg][]{waxman07,chevalier08,nakar10}.
For the case of blue supergiants or Wolf-Rayet stars,
these timescale are even shorter.
Other possible short-timescale transients
include, for example, the disk outflow
from black hole forming SNe
\citep[$<$ a few days,][]{kashiyama15}
and accretion induced collapse of white dwarfs \citep[$\sim$ 1 day,][]{metzger09}.
In addition to these,
there might also be unknown kind of transients with a short duration
since our knowledge on the short-timescale transients is still limited.

To explore the short-timescale transient sky,
some dedicated high-cadence surveys have started.
For example, Kiso Supernova Survey (KISS, \citealt{morokuma14,tanaka14agn},
using 1.05m Schmidt telescope and $\sim 4$ deg$^{2}$ wide field camera,
\citealt{sako12})
and High-cadence Transient Survey (HiTS, \citealt{forster14},
using 4m Blanco telescope and $\sim 3$ deg$^2$
Dark Energy Camera, \citealt{flaugher15})
adopt $\sim$ 1 hr cadence aiming at the detection of SN shock breakout.
There are also some ambitious surveys to explore even shorter timescales
\citep[\eg][]{becker04,rau08,berger13fot},
although no extragalactic transients with $\lsim$ 30 min timescale have been detected.

Recently, we have started a high-cadence transient survey
with the 8.2m Subaru telescope
and 1.77 deg$^2$ Hyper Suprime-Cam \citep[HSC,][]{miyazaki06,miyazaki12},
as a part of Subaru HSC Survey Optimized for Optical Transients (SHOOT).
SHOOT also adopts $\sim$ 1 hr cadence focusing on the detection of SN
shock breakout \citep{tominaga15}.
In this paper, we present rapidly rising transients discovered in SHOOT.
Here we define rapidly rising transients as objects
that rise more than 1 mag within restframe 1 day,
\ie the rising rate \dmdt\ > 1 mag day$^{-1}$.
We describe our observations and sample selection
in Section \ref{sec:observations}.
Then, we compare the obtained light curves
with previously known SNe and transients in Section \ref{sec:LC}.
Rising rates of various types of transients are summarized in Section
\ref{sec:dmdt}.
Based on these comparison,
we discuss the nature of these transients in Section \ref{sec:discussion}.
Finally we give conclusions in Section \ref{sec:conclusions}.
Throughout the paper, we assume the following cosmological parameters:
$\Omega_M = 0.273$, $\Omega_{\Lambda}= 0.726$,
and $H_0 = 70.5$ km s$^{-1}$ Mpc$^{-1}$ \citep{komatsu09}.
All the magnitudes are given in AB magnitude.

\begin{deluxetable}{lllll} 
\tablewidth{0pt}
\tablecaption{Log of observations}
\tablehead{
  UT     &  Epoch     & Instrument  & mode &  seeing$^{a}$ \\
         &            &             &       & (arcsec)
}
\startdata
2014-07-02 & Day 1  & HSC    & imaging ($g$,$r$) & 0.5\\
2014-07-03 & Day 2  & HSC    & imaging ($g$,$r$) & 0.6\\
2014-08-05 & Day 35 & FOCAS  & imaging ($g$,$r$) & 0.9 \\
            &       &         &  spectroscopy    &     \\
2014-08-06 & Day 36 & FOCAS  & imaging ($g$,$r$)  & 0.9 \\
            &       &         &  spectroscopy    &     \\
2015-05-24 & Day 327 & HSC$^{b}$  & imaging ($g$,$r$)  & 1.0 \\
2015-06-22 & Day 356 & FOCAS  & spectroscopy & 0.5 \\
2015-08-19 & Day 414 & HSC$^{b}$  & imaging ($r$)  & 1.4 \\
\enddata
\tablecomments{
  $^a$ Full width at half maximum.
  $^b$ Used as reference images.
}
\label{tab:log}
\end{deluxetable}

\section{Observations and sample selection}
\label{sec:observations}

\subsection{HSC observations}
\label{sec:HSC}

We performed a high-cadence transient survey with Subaru/HSC
for two continuous nights, 2014 July 2 and 3 UT
(hereafter Day 1 and 2, respectively).
The log of our observations is given in Table \ref{tab:log}.
Seven field-of-views ($\simeq 12\ {\rm deg^2}$) were
repeatedly visited with about 1 hr cadence.
Our survey was carried out mostly in optical $g$-band,
targeting the detection of the very early phase of SNe
\citep{tominaga15}.
Within one night, we had 3 or 4 visits in $g$-band
(here one ``visit'' consists of five 2-min exposures).
We also took 1 visit data in $r$-band in each night
to obtain $g-r$ color.

The HSC data were reduced using the HSC pipeline (version 3.6.1)
developed based on the LSST pipeline \citep{ivezic08,axelrod10}.
After standard reduction for each frame,
5 exposure images were co-added.
For astrometry and photometric calibration,
we used the Sloan Digital Sky Survey DR8 catalog \citep{aihara11}.
For stacked images for 1 visit (\ie 10 min exposure),
a typical limiting magnitude is about 26 mag
(5 sigma limiting magnitude for point sources)
in both $g$- and $r$-bands.

We performed image subtraction using the HSC pipeline.
The pipeline adopts the algorithm
developed by \citet{alard98} and \citet{alard00},
which are used for the ISIS package\footnote{\url{http://www2.iap.fr/users/alard/package.html}} and the HOTPANTS package\footnote{\url{http://www.astro.washington.edu/users/becker/v2.0/hotpants.html}}.
The algorithm uses a space-varying convolution kernel
to match the PSFs of two images.
The optimal convolution kernel is derived 
by minimizing the difference between 
convoluted PSFs of two images.
Although our 7 survey fields are selected 
based on the availability of the past imaging data,
most of the survey fields lack imaging data that
are deep and wide enough to be used as references for our new HSC images.
Thus, we used the data taken at the first visit of Day 1
as reference images for sample selection.

The data reduction described above was performed in realtime
using the on-site data analysis system \citep{furusawa11}
and a dedicated transient system \citep{tominaga15}.
By using these systems, transient candidates were
typically selected within the same night
\citep{tominaga14atel1,tominaga14atel2,tominaga15atel1,tominaga15atel2}.

To obtain the final reference images,
we also performed HSC imaging observations
on 2015 May 24 UT (Day 327, for $g$- and $r$-band)
and 2015 Aug 19 UT (Day 414, for $r$-band).
All the photometric values given in this paper are
derived by aperture photometry with 7 pixel radius (1.18 arcsec)
in the difference images using these final reference images.

\begin{deluxetable}{cccc} 
\tablewidth{0pt}
\tablecaption{Classification of detected sources}
\tablehead{
\multicolumn{4}{c}{Classification} \\
\multicolumn{4}{c}{Number of sources} 
}
\startdata
\multicolumn{4}{c}{Total} \\ 
\multicolumn{4}{c}{412} \\
\hline
\multicolumn{1}{c|}{Fake$^a$} &
\multicolumn{3}{c}{Astronomical objects} \\ 
\multicolumn{1}{c|}{215} &
\multicolumn{3}{c}{197} \\
\hline
\multicolumn{1}{c|}{} &
\multicolumn{1}{c|}{Star/quasar$^b$} &
\multicolumn{2}{c}{Non-star} \\
\multicolumn{1}{c|}{} &
\multicolumn{1}{c|}{166} &
\multicolumn{2}{c}{31} \\
\hline
\multicolumn{1}{c|}{} &
\multicolumn{1}{c|}{} &
\multicolumn{1}{c|}{Center} &
\multicolumn{1}{c}{Offset} \\
\multicolumn{1}{c|}{} &
\multicolumn{1}{c|}{} &
\multicolumn{1}{c|}{16 (8$^c$)} &
\multicolumn{1}{c}{15 (1$^c$)} \\ \hline
\enddata
\tablecomments{
  $^a$ Non-astronomical sources such as bad image subtraction, bad reference,
  or cosmic ray events. 
  $^b$ Point sources (including moving objects with a negligible motion).
  $^c$ Number of declining objects in the samples.
}
\label{tab:classification}
\end{deluxetable}

\begin{figure*}
\begin{center}
\begin{tabular}{ccccc}
\includegraphics[scale=0.9]{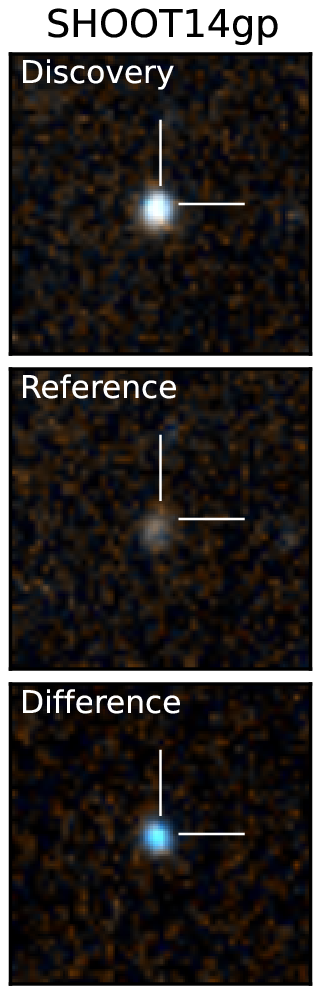} &
\includegraphics[scale=0.9]{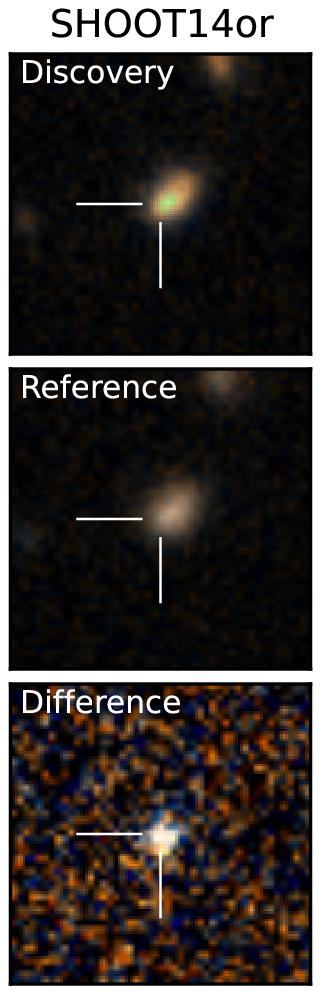} &
\includegraphics[scale=0.9]{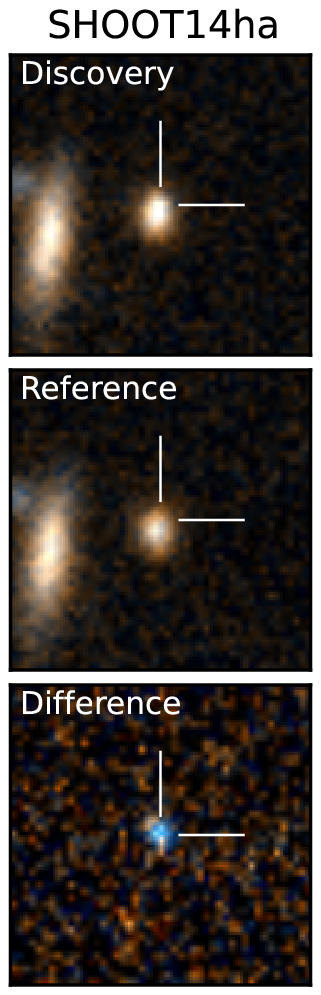} &
\includegraphics[scale=0.9]{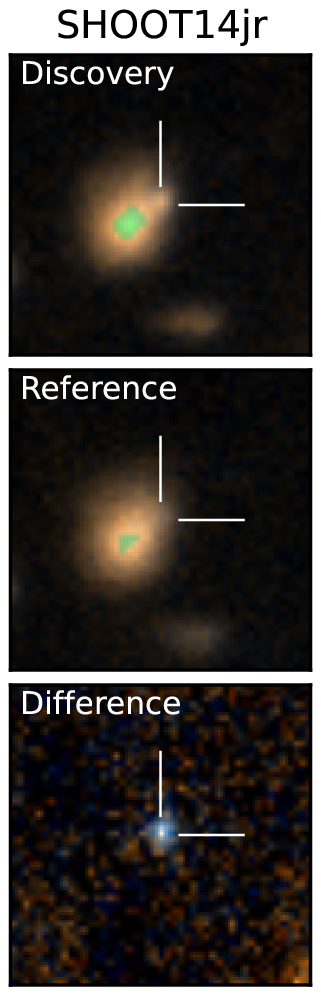} &
\includegraphics[scale=0.9]{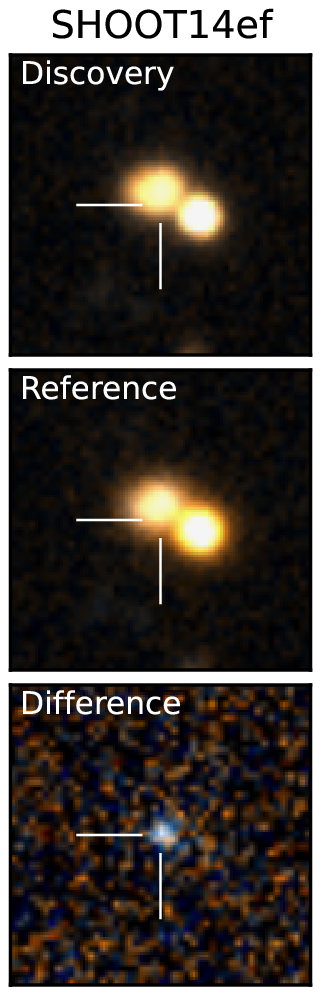}
\end{tabular}
\caption{
  Images of rapidly rising transients
  ($g$- and $r$-band two-color composite images).
  From top to bottom, each panel shows the discovery images taken on Day 2,
  images taken on Day 1 (used as references for the sample selection),
  and difference images (Day 2 $-$ Day 1).
  Each panel has $8'' \times 8''$ size. North is up and east is left.
  The color scale for the discovery and reference images are set to be the same.
}
\label{fig:image}
\end{center}
\end{figure*}

\begin{deluxetable*}{lcccc} 
\tablewidth{0pt}
\tablecaption{Rapidly rising transients from Subaru/HSC transient survey}
\tablehead{
Object &  R.A.     & Decl.     & Redshift &  \dmdt $^{a}$ \\
       & (J2000.0) & (J2000.0) &          &  (mag day$^{-1}$)
}
\startdata
SHOOT14gp  &  23:20:20.80  & +28:25:00.54    &  0.635   & $> 3.10$   \\
SHOOT14or  &  15:26:24.18  & +47:47:07.34    &  0.821   & $3.12^{+1.11}_{-0.70}$  \\
SHOOT14ha  &  23:21:44.91  & +28:54:49.80    &  0.548   & $>1.19$    \\
SHOOT14jr  &  16:33:49.99  & +34:28:05.36    &  0.384   & $1.61^{+0.39}_{-0.32}$  \\ 
SHOOT14ef  &  21:31:08.77  & +09:32:54.10    &  0.560   & $>1.31$  \\
\enddata
\tablecomments{
  $^a$ Measured in $g$-band data.
  Errors represent 1$\sigma$.
  For the objects that are not detected in the difference images
  on Day 1 (Day 1 $-$ Day 327),
  $3 \sigma$ lower limits are given.
}
\label{tab:objects}
\end{deluxetable*}

\begin{deluxetable}{cccc} 
\tablewidth{0pt}
\tablecaption{Photometry of rapidly rising transients}
\tablehead{
   MJD    & Filter &  Magnitude$^a$   &  Instrument  
}
\startdata
\multicolumn{4}{c}{SHOOT14gp} \\ \hline 
56840.542 & $g$ & $>25.53$              & HSC \\
56840.577 & $g$ & $>25.57$              & HSC \\
56841.513 & $g$ & $23.74^{+0.09}_{-0.08}$  & HSC \\
56841.547 & $g$ & $23.72^{+0.08}_{-0.07}$  & HSC \\
56841.582 & $g$ & $23.70^{+0.08}_{-0.07}$  & HSC \\
56840.560 & $g$ & $>25.58$              & HSC$^{b}$ \\
56841.548 & $g$ & $23.71^{+0.07}_{-0.06}$  & HSC$^{b}$ \\
56874.475 & $g$ & $>25.45$              & FOCAS \\
56840.479 & $r$ & $>24.99$              & HSC \\
56841.456 & $r$ & $24.31^{+0.15}_{-0.13}$  &  HSC \\
56874.463 & $r$ & $25.51^{+0.63}_{-0.39}$  &  FOCAS \\ \hline 
\multicolumn{4}{c}{SHOOT14or} \\ \hline 
56840.287 & $g$ & $26.74^{+0.65}_{-0.40}$  & HSC \\
56840.332 & $g$ & $26.88^{+0.75}_{-0.44}$  & HSC \\
56841.283 & $g$ & $25.11^{+0.13}_{-0.12}$  & HSC \\
56841.326 & $g$ & $25.01^{+0.12}_{-0.10}$  & HSC \\
56841.487 & $g$ & $24.99^{+0.11}_{-0.10}$  & HSC \\
56840.310 & $g$ & $26.85^{+0.64}_{-0.40}$  & HSC$^{b}$ \\
56841.365 & $g$ & $25.04^{+0.08}_{-0.07}$  & HSC$^{b}$ \\
56873.315 & $g$ & $>25.69$              & FOCAS \\
56840.431 & $r$ & $>25.61$             & HSC \\
56841.412 & $r$ & $25.25^{+0.25}_{-0.20}$ & HSC \\
56873.276 & $r$ & $25.79^{+0.59}_{-0.38}$ & FOCAS \\ \hline
\multicolumn{4}{c}{SHOOT14ha} \\ \hline 
56840.542 & $g$ & $>25.77$             &  HSC \\
56840.577 & $g$ & $>25.79$             &  HSC \\
56841.513 & $g$ & $25.29^{+0.33}_{-0.25}$ & HSC \\
56841.547 & $g$ & $25.27^{+0.28}_{-0.22}$ & HSC \\
56841.582 & $g$ & $24.95^{+0.19}_{-0.16}$ & HSC \\
56840.560 & $g$ & $>25.87$             & HSC$^{b}$ \\
56841.548 & $g$ & $25.11^{+0.20}_{-0.17}$ &  HSC$^{b}$ \\
56874.601 & $g$ & $>25.42$             & FOCAS \\
56840.479 & $r$ & $>25.48$             & HSC \\
56840.479 & $r$ & $>25.48$             & HSC \\
56841.456 & $r$ & $25.26^{+0.32}_{-0.25}$ & HSC \\
56873.501 & $r$ & $>25.03$             & FOCAS \\
56874.589 & $r$ & $>25.07$             & FOCAS \\ \hline
\multicolumn{4}{c}{SHOOT14jr} \\ \hline 
56840.299 & $g$ & $25.85^{+0.33}_{-0.25}$ & HSC \\
56840.342 & $g$ & $25.96^{+0.37}_{-0.27}$ & HSC \\
56840.526 & $g$ & $25.50^{+0.26}_{-0.21}$ & HSC \\
56841.293 & $g$ & $24.65^{+0.10}_{-0.09}$ & HSC \\
56841.338 & $g$ & $24.77^{+0.12}_{-0.11}$ & HSC \\
56841.500 & $g$ & $24.45^{+0.08}_{-0.08}$ & HSC \\
56840.389 & $g$ & $25.76^{+0.27}_{-0.21}$ & HSC$^{b}$ \\
56841.377 & $g$ & $24.61^{+0.09}_{-0.08}$ & HSC$^{b}$ \\
56840.442 & $r$ & $25.84^{+0.72}_{-0.43}$ & HSC \\
56841.422 & $r$ & $24.87^{+0.23}_{-0.19}$ & HSC \\
56873.262 & $r$ & $>25.36$             & FOCAS \\ \hline
\multicolumn{4}{c}{SHOOT14ef} \\ \hline 
56840.554 & $g$ & $>26.30$ & HSC \\
56840.591 & $g$ & $>26.41$ & HSC \\
56840.610 & $g$ & $>26.19$ & HSC \\
56841.525 & $g$ & $25.57^{+0.22}_{-0.18}$ & HSC \\
56841.559 & $g$ & $25.72^{+0.23}_{-0.19}$ & HSC \\
56841.596 & $g$ & $25.70^{+0.27}_{-0.21}$ & HSC \\
56841.615 & $g$ & $25.74^{+0.30}_{-0.23}$ & HSC \\
56840.585 & $g$ & $>26.50$             & HSC$^{b}$ \\
56841.574 & $g$ & $25.67^{+0.20}_{-0.17}$ & HSC$^{b}$ \\
56840.467 & $r$ & $>26.08$ & HSC \\ 
56841.445 & $r$ & $>26.06$ & HSC \\ 
\enddata
\tablecomments{
  $^a$ All the photometry are derived in the subtracted images
  using the final reference images.
  Errors represent 1$\sigma$.
  For the cases of non-detection, 3$\sigma$ upper limits are given.
  Magnitudes are corrected only for Galactic extinction.
  $^b$ Photometry in the 1-night stack images.
}
\label{tab:data}
\end{deluxetable}

\subsection{Sample selection}
\label{sec:selection}

We adopted the following selection processes
to select candidates for rapidly rising transients.
As mentioned above, we used the first images taken on Day 1
as reference images for the selection process.
Therefore, source detection in the subtracted images
is sensitive only to objects showing
variability within 2 nights.

Detected sources in the subtracted images contain
not only real astronomical sources
but also fake sources such as spikes around bright stars,
and artifacts due to mis-subtraction or mis-alignment
\citep[\eg][]{bailey07,bloom12,brink13}.
Thus, we selected objects detected in the subtracted images
at least twice with $>5\sigma$ significance.
After this selection, 1407 sources remain.
We first performed initial visual screening,
resulting in 430 sources with SHOOT14XX names
(412 independent sources because of 18 duplication
in overlapped regions in the reduced images).
Then, we further performed detailed classification.
Results of the classifications are summarized
in Table \ref{tab:classification}.

Among 412 independent sources,
215 sources are still fakes of the subtracted images
while the other 197 sources are likely to be astronomical sources.
The astronomical sources are dominated by
stellar-shape sources, such as stars or quasars (166 sources).
The remaining 31 sources are associated with extended sources (galaxies).
Among these sources, 16 sources are located at the center of galaxies.
Since they may be active galactic nuclei or tidal disruption events,
we avoided these objects for follow-up observations.
Since 8 out of 16 objects show declining flux,
it is likely that the majority of these 16 sources are active galactic nuclei.
Remaining 15 sources have an offset from the center of the galaxies,
and selected as SN candidates.

The final SN candidates consist of 14 brightening objects.
From this final sample, we performed follow-up observations
of most reliable 12 objects.
Among these 12 objects, we measured redshifts for 8 objects
while the other 4 objects (and their host galaxies) were
too faint to take spectra.
The remaining 2 objects were not observed.

Note that the sample selection for spectroscopy was made based
on the {\it flux} difference within 2 nights,
not on the {\it magnitude} difference
since the final reference images were not available
and true magnitudes of the objects on Day 1 were not known
at the time of spectroscopy (2014 Aug).
Therefore, even after the selection processes,
our initial samples could include not only rapidly rising transients
but also normal SNe around the peak brightness
if the flux difference within 2 nights is large enough.
In fact, by our follow-up spectroscopic observations
(Section \ref{sec:followup}),
3 out of 8 objects were identified as normal SNe
(at $z$=0.13, 0.25, and 0.40).
In addition, after obtaining the final reference images on Day 327,
we confirmed that these three objects are already bright on Day 1.
The rising rates for these three objects are \dmdt $< 1$ mag day$^{-1}$,
which is also consistent with normal SNe.
Therefore, we omit these three objects from our samples.

Figure \ref{fig:image} shows images of
5 rapidly rising transients,
named as SHOOT14gp, 14or, 14ha, 14jr, and 14ef
(Table \ref{tab:objects}).
Photometry of these 5 objects is shown in Table \ref{tab:data}.

\begin{figure}
\begin{center}
\includegraphics[scale=1.0]{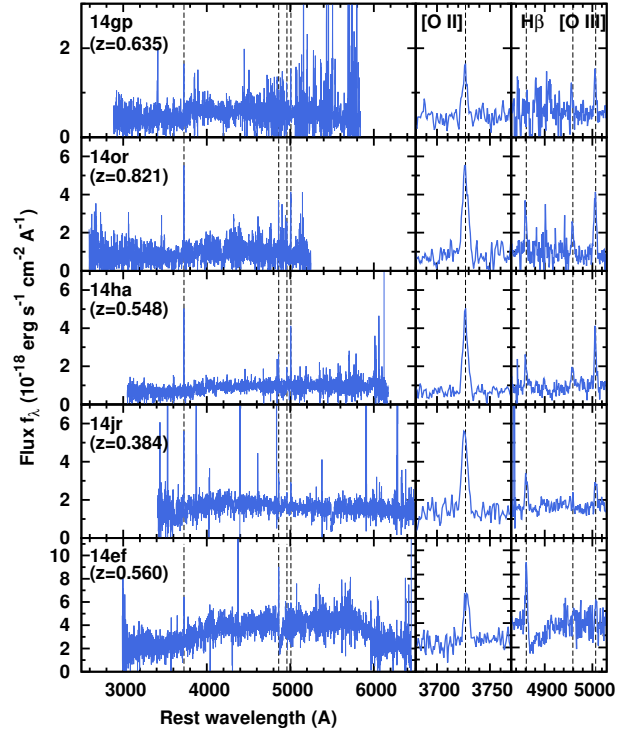} 
  \caption{
      Spectra of host galaxies of SHOOT14gp, 14or, 14ha, 14jr, and 14ef
      (from top to bottom).
      The wavelengths of strong emission lines
      ([\ion{O}{ii}] $\lambda$3727, H$\beta$, and [\ion{O}{iii}] $\lambda$4959,5007) are marked with
      the dashed lines.
      The right panels show the data around these lines.
}
\label{fig:spec}
\end{center}
\end{figure}

\subsection{Follow-up observations}
\label{sec:followup}

We performed imaging and spectroscopic observations of
5 objects (Table \ref{tab:objects})
using the Faint Object Camera and Spectrograph
\citep[FOCAS,][]{kashikawa02} of the Subaru telescope.
Observations of the four objects
(SHOOT14gp, 14or, 14ha, and 14jr)
were carried out on 2014 Aug 5 and 6 UT (Day 35 and 36, respectively)
while observations of SHOOT14ef were on 2015 June 22
(Day 356, only for the host galaxy).

For the FOCAS imaging data, we performed image subtraction
with the final reference images using
HOTPANTS package.
SHOOT14gp and 14or were marginally detected only in $r$-band
while they were not detected in $g$-band.
The other objects were not detected both in $g$- and $r$-bands.
A typical limiting magnitudes are $\simeq 25.0-25.5$ mag
(Table \ref{tab:data}).

For spectroscopy, we used multi-object mode with $0\farcs 8$-width slit
and long-slit mode with $1\farcs 0$-width slit (only for SHOOT14ef).
With the 300B (300 lines mm$^{-1})$ grism and the SY47 order-sort filter,
our configuration gives a wavelength coverage of 4700 - 9000 \AA\
and a spectral resolution of $R = \lambda/\Delta \lambda \sim 600$.
The data were reduced with the IRAF packages in a standard manner.

The transient components are not detected in our spectra
as expected from the results of imaging observations.
Figure \ref{fig:spec} shows the spectra of the host galaxies
for these five objects.
The [\ion{O}{ii}] $\lambda$3727 emission line is detected from all the host galaxies,
which indicates that they are all star forming galaxies.
The redshifts range from $z =0.384$ (SHOOT14jr) to $z=0.821$ (SHOOT14or).

\begin{figure*}
\begin{center}
\begin{tabular}{ccccc}
  \includegraphics[scale=1.1]{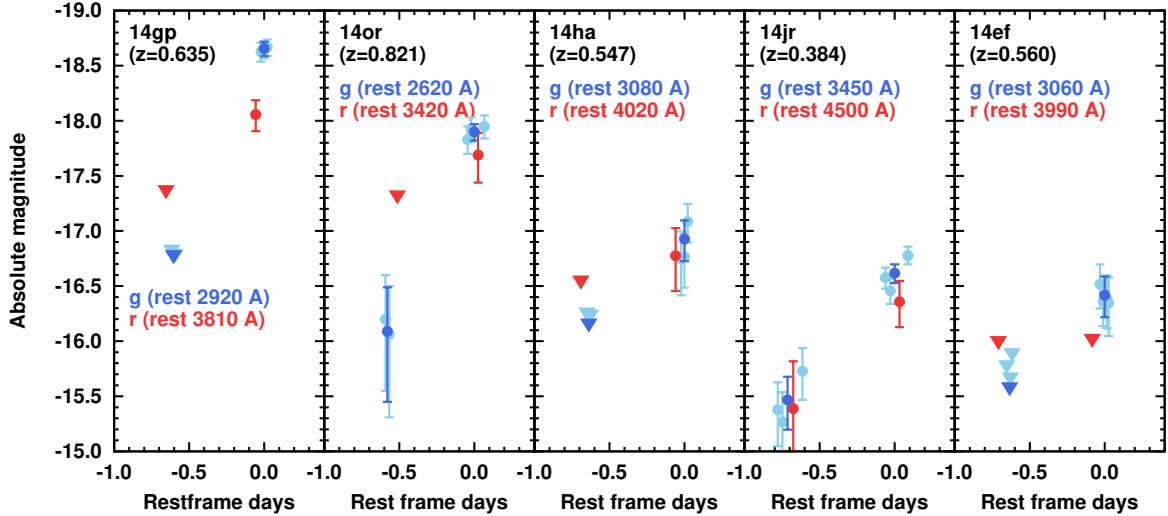} 
\end{tabular}
\caption{Light curves of the five rapidly rising transients on Days 1 and 2.
  The $g$- and $r$-band photometry is shown in blue and red points.
  Triangles show $3 \sigma$ upper limit.
  For the $g$-band data, photometry for 1 visit
  (5 $\times$ 2-min exposures) is shown in pale blue color
  while photometry in the 1-night stacked data is shown in blue color.
}
\label{fig:LCabs}
\end{center}
\end{figure*}

\section{Light curves}
\label{sec:LC}

\subsection{Overview}
\label{sec:overview}

Figure \ref{fig:LCabs} shows light curves of
our samples on Day 1 and Day 2.
Hereafter, the epochs of stacked $g$-band data on Day 2 are taken
to be $t=0$ unless otherwise mentioned.
The photometry is performed in the subtracted images
using the final references
(\eg Day 1 $-$ Day 327 and Day 2 $-$ Day 327 for $g$-band).

Throughout the paper,
we do not take into account full $K$-correction for absolute magnitudes
since only limited information about
spectral energy distribution is available for our samples.
Instead, we only correct the effect of redshifts, \ie
$M = m - \mu + 2.5\log(1+z)$, where $M$ and $m$ are absolute
and observed AB magnitudes (measured as $f_{\nu}$),
$\mu$ is the distance modulus.
The last term originates from the difference in the frequency bin 
in the restframe and observer frame, \ie
$L_{\nu}(\nu_e) = [(4 \pi d^2)/(1+z)] f_{\nu}(\nu_o)$,
where $\nu_e$ and $\nu_o$ are restframe and observer frame frequency,
and $d$ is the luminosity distance \citep{hogg02}.

The absolute magnitudes of the five objects
range from $-16$ to $-19$ mag in the restframe near-ultraviolet (UV)
wavelengths (2620\AA $-$ 3450\AA, depending on the redshifts).
The photometric values of our samples
are corrected for the extinction in our Galaxy
but not for the extinction in the host galaxy.
Therefore, intrinsic absolute magnitudes can be brighter
than those shown in Figure \ref{fig:LCabs}.

All of the five objects show blue $g-r$ color on Day 2,
$g-r \simeq -0.60, -0.21, -0.15$, and $-0.15$ mag for
SHOOT14gp, 14or, 14ha, and 14jr, respectively.
For SHOOT14ef, the color is $g-r < -0.39$ mag.
This indicates that, for blackbody case, the peak of the spectra
is located at wavelengths shorter than
the wavelengths corresponding to the observed $r$-band.
Therefore the blackbody temperatures for our objects are
$T_{\rm BB} \gsim$ 13000, 15000, 13000, 11000, and 13000 K for
SHOOT14gp, 14or, 14ha, 14jr, and 14ef, respectively.
Note that the intrinsic colors can be bluer
due to the extinction in the host galaxies.

SHOOT14or and 14jr are detected in the
images of Day 1 $-$ Day 327.
We measure the rising rates from Day 1 to Day 2
using the $g$-band 1-day stacked images:
\dmdt = $3.12^{+1.11}_{-0.70}$ and $1.61^{+0.39}_{-0.32}$ mag day$^{-1}$
for SHOOT14or and 14jr, respectively
(errors represent 1$\sigma$, Table \ref{tab:objects}).
Note that the rising rate is measured in the restframe,
so the time interval used for the measurement varies with
the source redshifts ($\Delta t = 0.55$ days for SHOOT14or
while $\Delta t = 0.72$ days for SHOOT14jr).
The other three objects (SHOOT14gp, 14ha, and 14ef) are not detected
in the subtracted images of Day 1 $-$ Day 327.
The 3 $\sigma$ lower limits of the rising rate measured in $g$-band
are \dmdt \ $>$ 3.10, 1.21, and 1.17 mag day$^{-1}$.
These are also high enough to match our criterion for rapidly rising
transients.

In the following sections,
we compare the light curves of our samples
with those of previously known SNe and transients.

\begin{figure*}
\begin{center}
\begin{tabular}{cc}
  \includegraphics[scale=0.85]{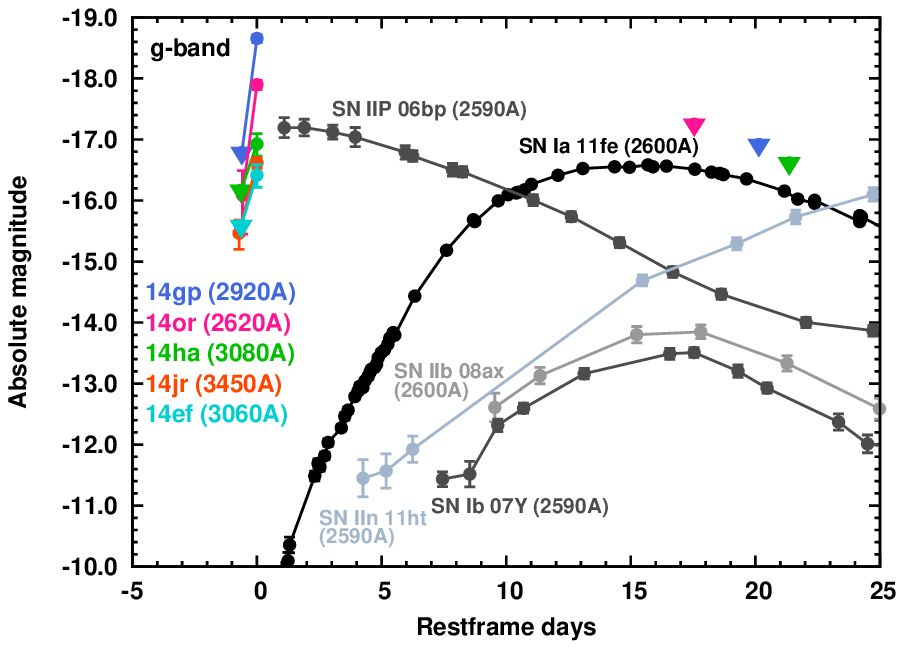} &
  \includegraphics[scale=0.85]{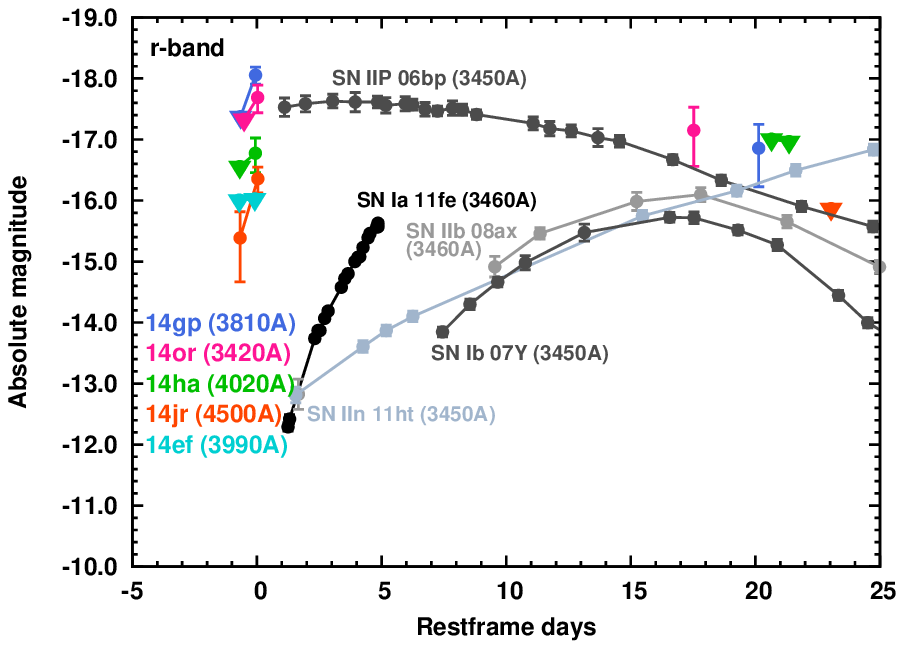} 
\end{tabular}
    \caption{
    {\it Left}: Comparison between $g$-band light curves of our objects 
    and {\it Swift} $uvw1$-band light curves of nearby normal SNe:
    Type Ia SN 2011fe \citep{brown12}, Type IIP SN 2006bp,
    Type IIb SN 2011dh, Type IIn SN 2011ht,
    and Type Ib SN 2007Y \citep{pritchard14}.
    {\it Right}: Comparison between $r$-band light curves of our objects and
    {\it Swift} $u$-band light curves.
    For {\it Swift} SN data, the estimated epoch of the explosion
    is taken to be $t=0$ day.
    The {\it Swift} data are corrected for the extinction
      both in our Galaxy and host galaxies
      as estimated by \citet{pritchard14}.
    Vega magnitudes are converted to AB magnitudes
    using the zeropoints presented by \citet{breeveld11}.
}
\label{fig:LCall}
\end{center}
\end{figure*}

\begin{figure}
\begin{center}
  \includegraphics[scale=0.9]{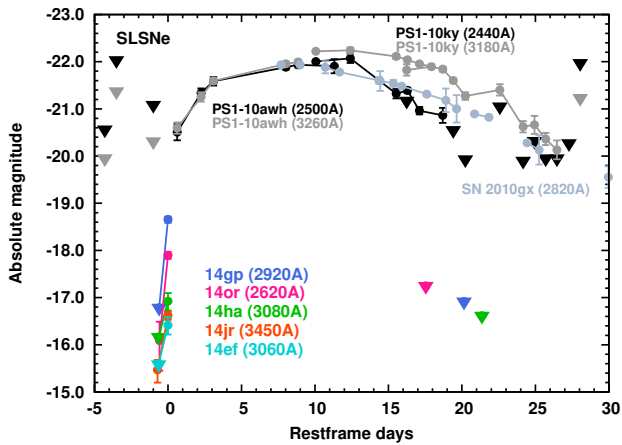}
\caption{Comparison of light curves
  with SLSNe \citep{pastorello10,chomiuk11}.
  Observed $u$-band light curves are shown for SN 2010gx,
  while observed $g$- and $r$-band light curves are shown for PS1-10awh
  and PS1-10ky.
  For SLSNe, the peak epochs are shifted to $t=13$ days
  and magnitudes are corrected for only Galactic extinction.
}
\label{fig:LCSLSN}
\end{center}
\end{figure}

\subsection{Comparison with SNe}
\label{sec:SN}

Figure \ref{fig:LCall} shows comparison of rapidly rising transients
with normal SNe.
Since the redshifts of our samples are moderately high,
$z =0.384-0.821$,
we compare our $g$- and $r$-band light curves with
near-UV and $u$-band light curves
of nearby SNe with good temporal coverage.
We use the {\it Swift} $uvw1$- and $u$-band data
from \citet{brown12} and \citet{pritchard14} with
extinction correction (both in our Galaxy and host galaxies)
using the extinction law by \citet{brown10}.
Since the effective restframe wavelengths do not always match perfectly,
we always give effective restframe wavelengths in parenthesis.

Figure \ref{fig:LCall} shows that 
the properties of our samples are not
consistent with those of Type Ia SNe at any phase, and
those of core-collapse SNe at $\gsim$ a few days after the explosion.
The absolute magnitudes of our samples are as luminous as the peak
magnitude of Type Ia SN 2011fe \citep{brown12}
and Type IIP SN 2006bp \citep{pritchard14}.
However, the rising rates for our samples are faster than
the very early phase of SN 2011fe, one of the best observed Type Ia SNe.
We also compare our objects with Type IIb SN 2008ax, Type IIn SN 2011ht,
and Type Ib SN 2007Y \citep{pritchard14}.
Their rising rates are slower than those of our samples at any epochs
with available data, \ie $\gsim$ a few days after the explosion.
In addition, the blue colors of our samples ($g-r \le -0.2$ mag)
are not consistent with normal SNe after a few days
from the explosion.
For nearby SNe after a few days from the explosion,
the $uvw1$ magnitude is generally fainter than
the $u$ magnitude as shown in Figure \ref{fig:LCall},
\ie the color is $uvw1-u > 0$ mag.

Our samples might correspond to the rising
phase of much brighter SNe, such as superluminous
SNe \citep[SLSNe, ][]{quimby11,gal-yam12}.
Figure \ref{fig:LCSLSN} shows comparison of our samples
with SLSN SN 2010gx, PS1-10awh, and PS1-10ky
with a good temporal coverage \citep{pastorello10,chomiuk11}.
Our data on Days 1 and 2 could be interpreted as
the very early phase of SLSNe, which have never been caught.
However, the data on Days 35 and 36 are clearly inconsistent
with the declining part of SLSNe.

\begin{figure*}
\begin{center}
\begin{tabular}{cc}
\includegraphics[scale=0.85]{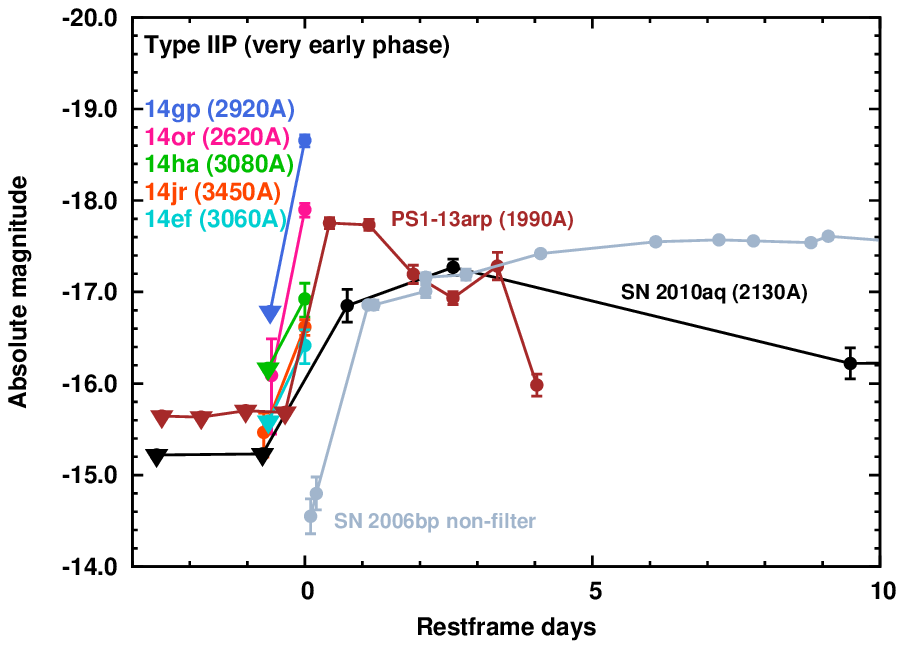} &
\includegraphics[scale=0.85]{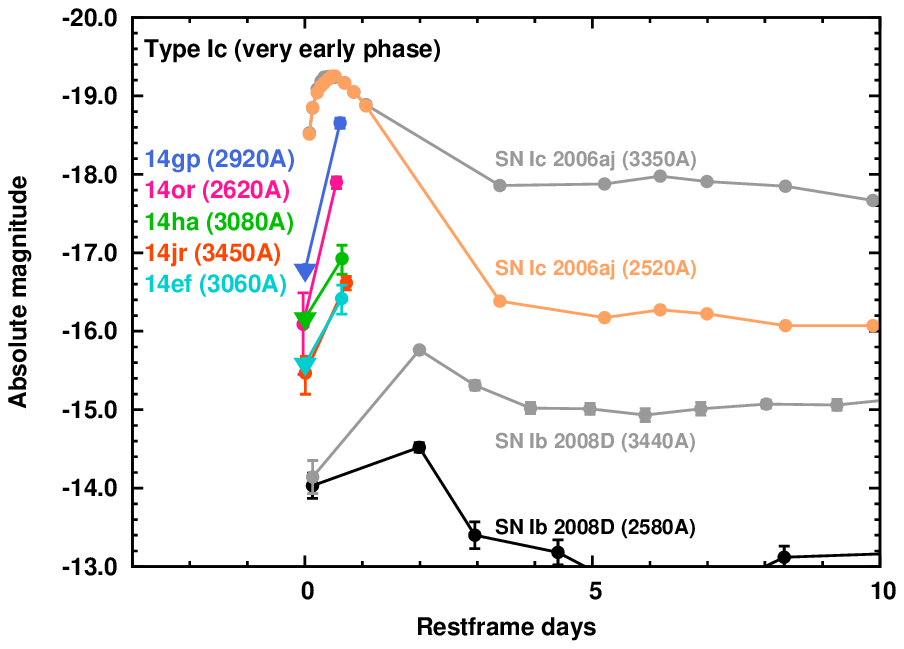}
\end{tabular}
\caption{
  {\it Upper}: Comparison of light curves with the very early phase of Type IIP SNe:
  {\it GALEX} NUV data of Type IIP SN 2010aq \citep{gezari10} and
  PS1-13arp \citep{gezari15},
  and also non-filter data of SN 2006bp \citep{quimby0706bp}.
  The data of SN 2010aq and PS1-13arp are corrected for only Galactic extinction
    while those of SN 2006bp are corrected for both Galactic and host extinction.
  {\it Lower}: Comparison with the very early phase of
  Type Ic SN 2006aj \citep[corrected for only Galactic extinction]{campana06,simon10}
  and Type Ib SN 2008D \citep[corrected for both Galactic and host extinction]{modjaz09}.
  For the comparison with Type Ibc SNe, the epoch of our data
  are shifted so that Day 1 corresponds to $t=0$ day.
}
\label{fig:LCearly}
\end{center}
\end{figure*}

\begin{figure*}
\begin{center}
\begin{tabular}{cc}
  \includegraphics[scale=0.85]{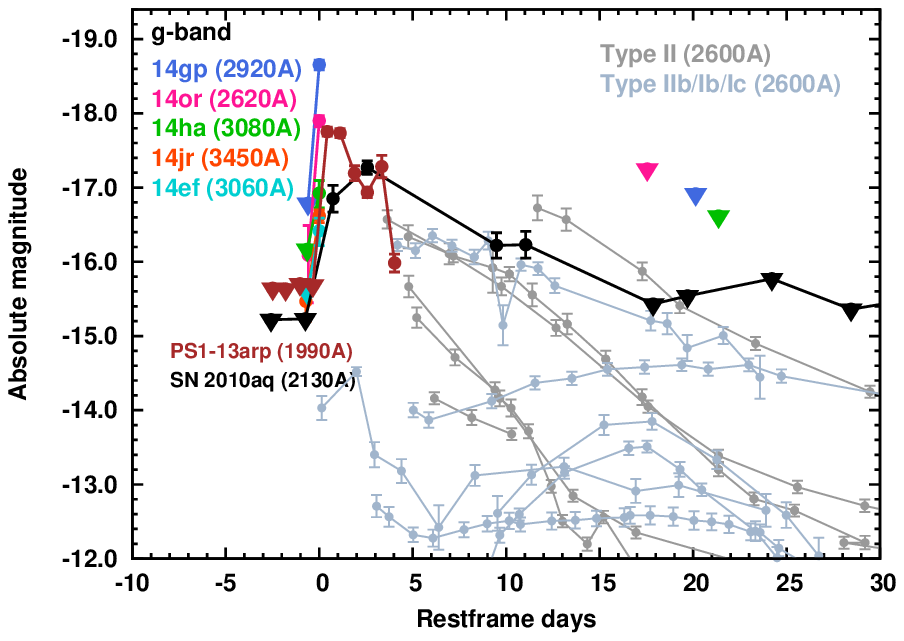}
  \includegraphics[scale=0.85]{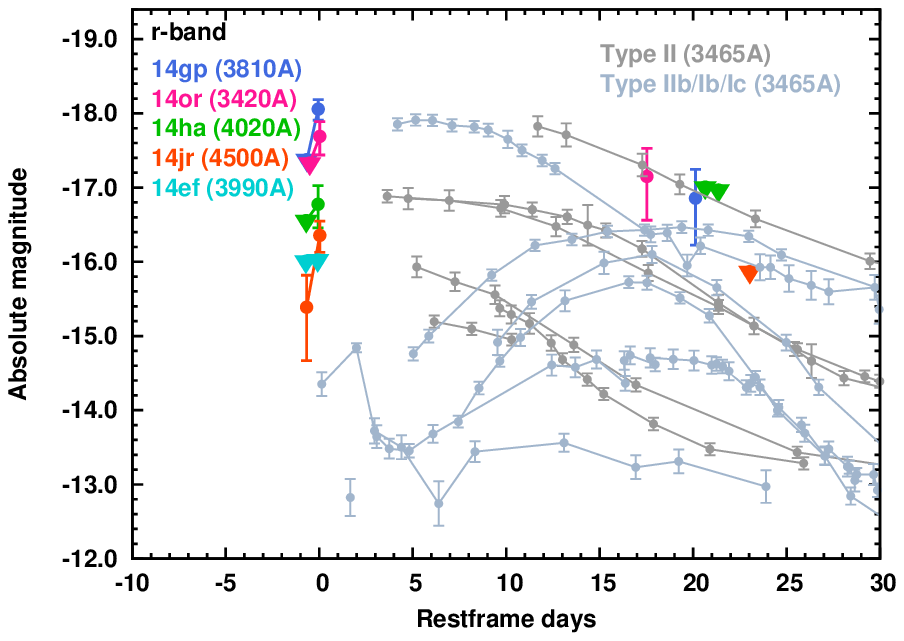} 
\end{tabular}
  \caption{
  {\it Left}: Comparison between $g$-band light curves of our objects
  and {\it Swift} $uvw1$-band light curves of core-collapse SNe
  \citep{pritchard14,modjaz09} and         
  Type IIP SN 2010aq \citep{gezari10} and
  PS1-13arp \citep{gezari15} with {\it GALEX} NUV data.
  {\it Right}: Comparison between $r$-band light curves of our objects
  and {\it Swift} $u$-band light curves of core-collapse SNe.
  The data from \citet{pritchard14} are corrected for
    estimated extinction
    both in our Galaxy and host galaxies.
  Vega magnitudes are converted to AB magnitudes.
}
\label{fig:LCCCSN}
\end{center}
\end{figure*}

\subsection{Comparison with very early phase of SNe}
\label{sec:earlyCC}

We compare our samples with earlier phases of SNe
($\lsim$ a few days after the explosion).
First, we show comparison with
Type IIP SN 2010aq \citep{gezari10} and PS1-13arp \citep{gezari15},
with UV detection at the very early phase with $GALEX$.
The early emission of SN 2010aq is consistent with 
cooling envelope emission after SN shock breakout \citep{gezari10}.
The emission of PS1-13arp is brighter and shorter,
which may indicate shock breakout emission from dense wind \citep{gezari15}.

The upper panel of Figure \ref{fig:LCearly} shows 
a similarity of the rising rate and brightness
between our samples and SN 2010aq and PS1-13arp.
SN 2010aq and PS1-13arp also show fast rise, \dmdt $ > 0.989$
and $> 2.635$ mag day$^{-1}$, respectively.
They reach about $-17$ - $-18$ mag, which is also similar to our samples.
Note that the effective restframe wavelengths corresponding
to the NUV filter of $GALEX$ (2130 \AA\ and 1990 \AA\ for
SN 2010aq and PS1-13arp, respectively)
are shorter than those for our samples ($\sim 2600-3500$ \AA).

For comparison, we also show non-filter magnitude of 
Type IIP SN 2006bp \citep{quimby0706bp},
for which very early phases were observed
(see also \citealt{rubin15} for recent larger samples).
It also shows a fast rise, \dmdt $= 2.3$ mag day$^{-1}$.
Again, although the difference in the restframe wavelengths
should be cautioned,
these similarities suggest that our samples of rapidly rising
transients are the very early phase of SNe.

We also compare our samples with the very early part of
Type Ic SN 2006aj and Type Ib SN 2008D.
They are among the best-studied stripped-envelope SNe.
SN 2006aj is associated with low luminosity gamma-ray burst (GRB) 060218,
and thus, good optical to NUV data are available
from soon after the explosion
\citep[\eg][]{campana06,soderberg06,pian06,mazzali06,sollerman06,modjaz06,mirabal06,simon10}.
SN 2008D is associated with X-ray transient 080109
\citep[\eg][]{soderberg08,mazzali08,tanaka0908D,tanaka0908Dneb,modjaz09}.
Emission at the first 2 days of SN 2006aj and SN 2008D is
interpreted as cooling envelope emission
\citep{waxman07,soderberg08,modjaz09,chevalier08,nakar15}.

The lower panel of Figure \ref{fig:LCearly} shows that
the rising rate of SN 2006aj is as fast as our samples.
The time to the peak is only $\sim 0.5$ days,
which is as short as that inferred for our samples
although we cannot not firmly determine the peak dates only with 2-night data.
SN 2008D lacks the data at $\sim 1$ day after the explosion.
Nevertheless, the rising rate of SN 2008D in {\it Swift} $u$-band
(measured with 2-day interval) is similar to SHOOT14jr.
Note that if the early part of SN 2008D is interpreted as
cooling envelope emission, the peak would be around $\sim 1$
day after the explosion \citep{soderberg08,modjaz09},
and the rising rate in the first day
is faster than that measured with 2-day interval.

When we match our objects with 
core-collapse SNe within a few days after the explosion,
our observations on Day 35 and 36 correspond to
the plateau phase of Type IIP or the peak phase of Type Ibc SNe.
As shown in Figure \ref{fig:LCCCSN}, the distribution of $uvw1$ brightness
of core-collapse SNe at these epochs ranges from $-12$ to $-17$ mag.
Since our limits in $g$-band correspond to $-17.0$ mag, 
non-detection in $g$-band on Days 35 and 36 is not surprising.
SHOOT14gp and 14or are marginally detected in $r$-band
(right panel of Figure \ref{fig:LCCCSN}).
Compared with {\it Swift} $u$-band data, their brightness is
consistent with those of core-collapse SNe at the luminous end.

\begin{figure*}
\begin{center}
  \begin{tabular}{cc}
    \includegraphics[scale=0.85]{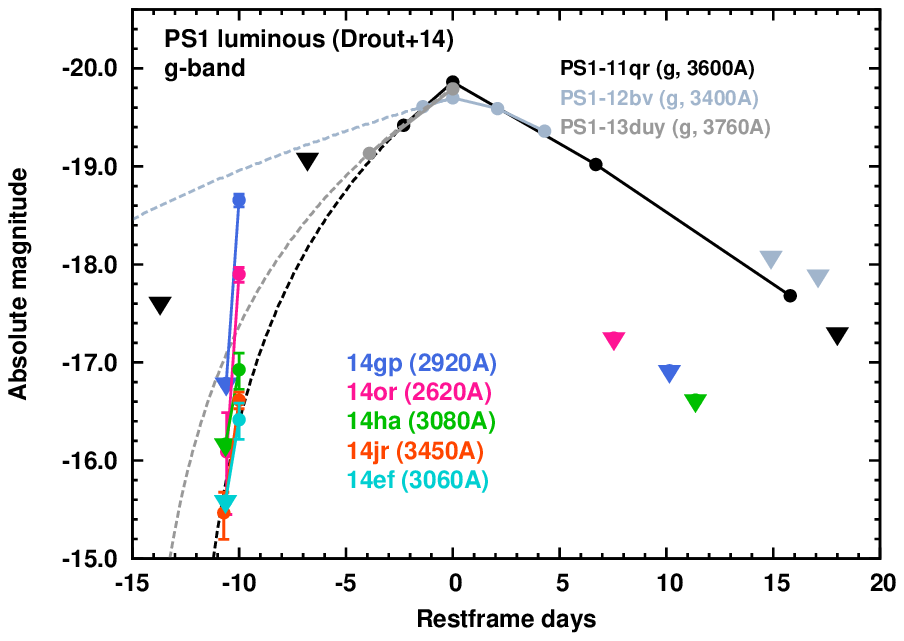} &
    \includegraphics[scale=0.85]{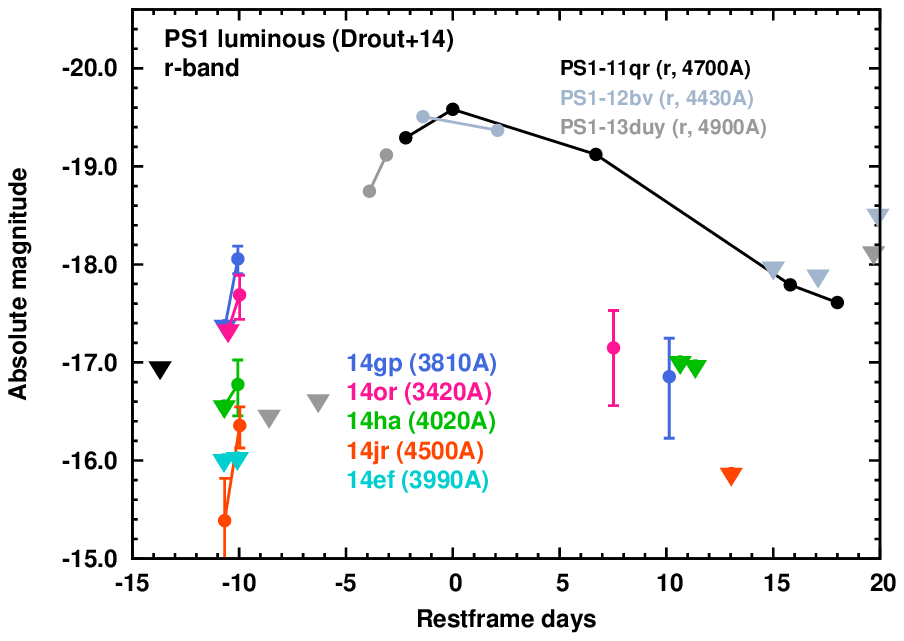} \\
    \includegraphics[scale=0.85]{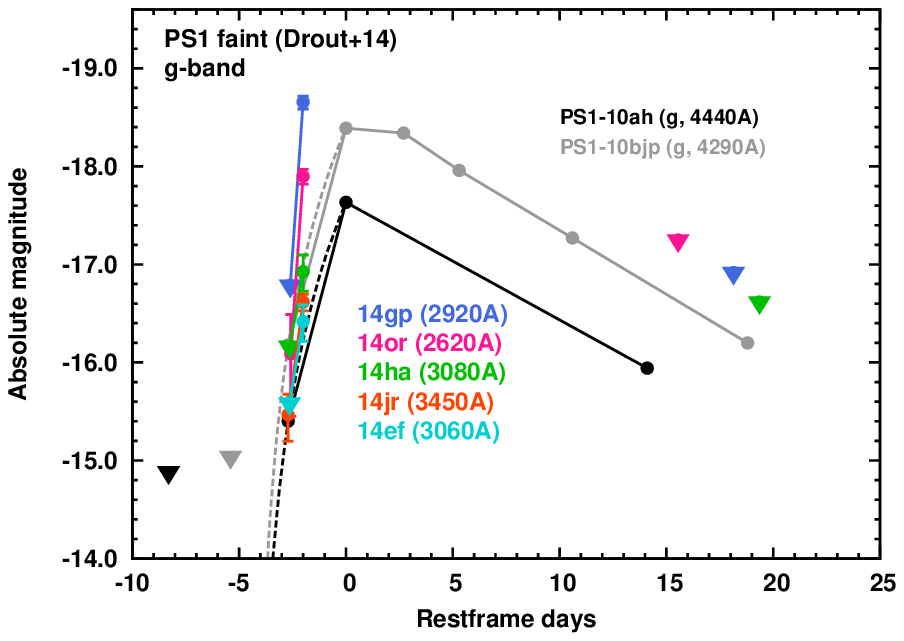} &
    \includegraphics[scale=0.85]{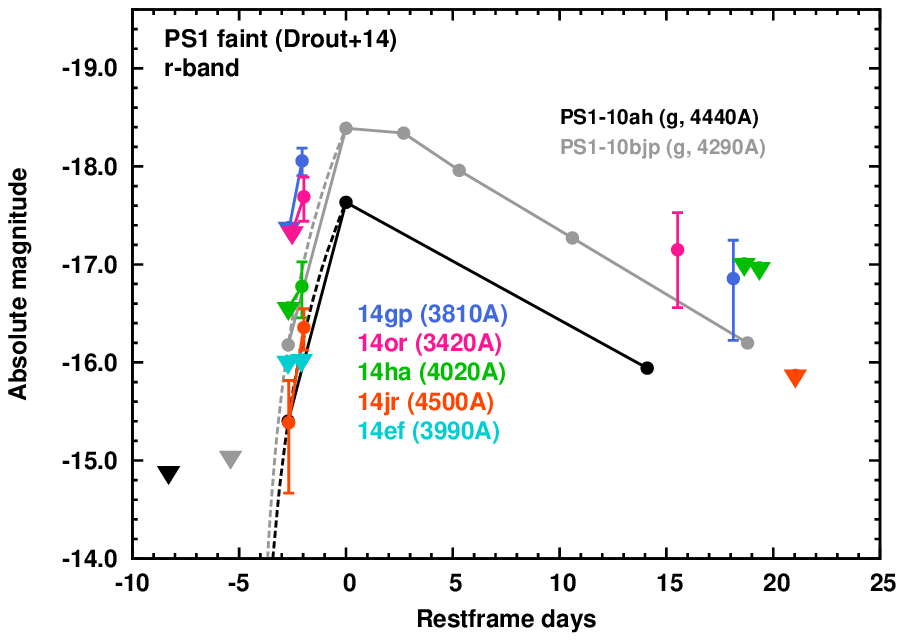} 
  \end{tabular}
    
  \caption{Comparison of light curves
    with rapidly evolving and luminous transients from PS1
    \citep{drout14}.
    The peak epoch of the PS1 samples is selected to be $t=0$ day.
    {\it Upper}: Comparison with the PS1 luminous samples with
    the peak absolute magnitudes of $< -19$ mag.
    Left and right panels show the light curves in $g$- and $r$-band
    (both for HSC and PS1), respectively.
    Epochs of our samples are shifted so that Day 2 data
    correspond to be $t= -10$ days.
    {\it Lower}: Comparison with the PS1 faint samples with
    the peak absolute magnitudes of $> -19$ mag.
    Left and right panels show the light curves in $g$- and $r$-band
    for HSC data, respectively.
    For the PS1 sample, $g$-band data are shown in the both panels
    (as $g$-band has closer effective wavelengths).
    Epochs of our samples are shifted so that Day 2 data
    correspond to be $t= -2$ days.
    The magnitudes of the PS1 samples are corrected for only Galactic extinction.
}
\label{fig:LCrapid}
\end{center}
\end{figure*}

\begin{figure*}
\begin{center}
    \includegraphics[scale=1.5]{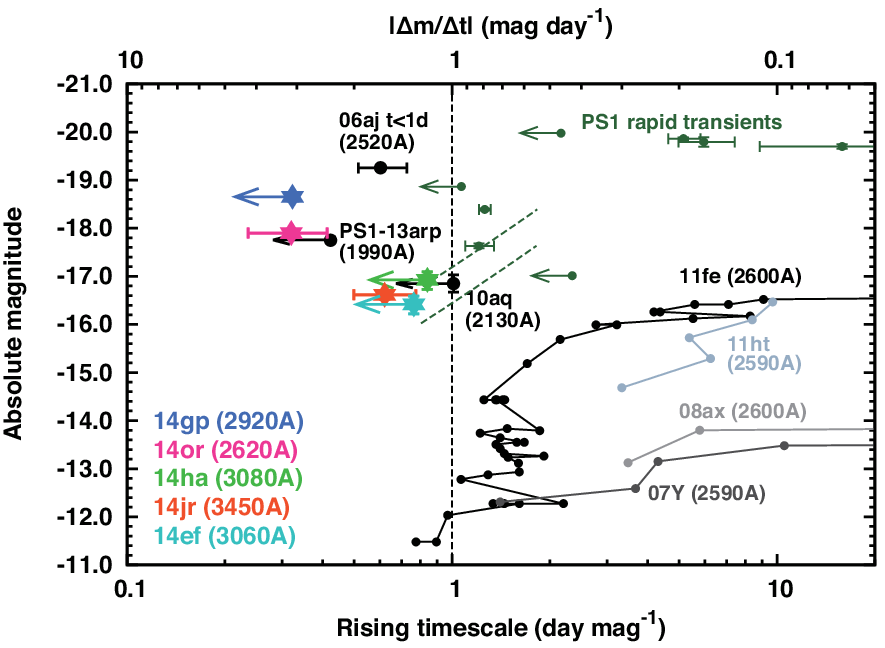}
\caption{Summary of absolute magnitudes and rising timescale
  ($\tau_{\rm rise} \equiv 1/$ \dmdt) of transients.
  Our samples are compared with the following objects:
  SN 2010aq and PS1-13arp \citep{gezari10,gezari15}
  with early UV detection with {\it GALEX},
  the early peak of SN 2006aj
  \citep[,Figure \ref{fig:LCearly}]{campana06,simon10},
  Type Ia SN 2011fe \citep{brown12}, core-collapse SNe
  (Type Ib SN 2007Y, Type IIb SN 2008ax, and Type IIn SN 2011ht,
  \citealt{pritchard14}),
  and rapid transients from PS1 \citep{drout14}.
  For rapid transients from PS1, the rising timescale (rising rate) is
  measured with $g$-band data.
  The dashed lines show the absolute magnitude and rising timescale
  of PS1-10ah and PS1-10bjp measured with the interpolated $g$-band light curves.
}
\label{fig:dmdt}
\end{center}
\end{figure*}

\subsection{Comparison with rapidly rising transients from PS1}
\label{sec:rapid}

The rapid rising rates of our samples remind us of
population of rapidly evolving and luminous transients
from PS1, which are compiled by
\citet[see also \citealt{poznanski10,kasliwal10,drout13}]{drout14}.
These transients show rapid luminosity evolution
both in rising and declining phases compared with normal SNe
with a time above half-maximum of less than 12 days.
Interestingly, they show a faster rising rate than a declining rate,
which motivates the comparison with our samples.
In addition, they have blue $g-r$ colors ($g-r < -0.2$ mag),
similar to our samples.

Since the PS1 samples have a wide luminosity range,
we divide the samples into two classes with
the absolute magnitude brighter (hereafter PS1 luminous samples) or
fainter (PS1 faint samples) than $-19.0$ mag.
\citet{drout14} interpret their rapid transients to be either
(1) the cooling envelope emission following shock breakout
(especially for faint samples)
or (2) shock breakout from dense wind (for luminous samples).

Figure \ref{fig:LCrapid} shows comparison of our samples with
the PS1 samples \citep{drout14} which are detected at the rising part in $g$-band.
The peak dates of the PS1 samples are taken to be $t=0$ day.
It should be cautioned that the PS1 samples have
a wider redshift range than ours,
and thus the rest wavelengths corresponding the observed filters
have a wider variety.
For the PS1 luminous samples, $g$- and $r$-band data for our samples
are compared with PS1 $g$- and $r$-band data, respectively.
Since the PS1 faint samples have low redshifts ($z = 0.074$ and 
$0.113$ for PS1-10ah and PS1-10bjp, respectively),
we compare our $g$- and $r$-band data with PS1 $g$-band data.

The peak magnitudes of the PS1 luminous samples are brighter
than the magnitudes of our sample on Day 2.
Our samples could thus be interpreted to the rising part of the PS1 samples.
The dashed lines in the upper left panel of Figure \ref{fig:LCrapid}
shows the extrapolation of the rising part
by assuming the flux rises as $f = (t-t_0)^2$
(as often assumed for the early part of SNe,
see \eg \citealt{nugent11,pastorello13,prieto13,yamanaka14}),
where $t_0$ is the epoch with zero flux.
Three of our samples (SHOOT14ha, 14jr, and 14ef) show a nice agreement
with the extrapolated rising part
if the epochs of these objects are shifted
so that Day 2 corresponds to $t \sim -10$ days.
However, with this assumption, the non detection of PS1-13duy
before the peak in $r$-band is not consistent with our detection on Day 2.
In addition, the brightness and upper limits at later epochs
(Days 35 and 36) are much fainter than the magnitudes of PS1-11qr
for which the data at the declining part is available.
Therefore, our samples are not likely to be the same population
as the PS1 luminous samples.

Our samples show a better agreement with the PS1 faint samples
(lower panels of Figure \ref{fig:LCrapid}).
The rising rates of the PS1 samples in $g$-band is
\dmdt\ < 1 mag day$^{-1}$, which do not fulfill our criterion.
However, PS1 data are taken with $\sim 3$ days cadence,
and thus, the rising rate measured with a shorter interval can be faster.
In fact, if the rising part is
interpolated with $f = (t-t_0)^2$,
the rising rate can be as fast as that measured for our samples.
Especially, three of our samples (SHOOT14ha, 14jr, and 14ef) show
a good match if the epochs of these objects are shifted
so that Day 2 corresponds to $t \sim -2$ days.
Then, our data at later epochs are also consistent with
the PS1 samples at the declining phase.
Since the estimated epoch of zero flux for PS1-10ah and PS1-10bjp
is $t_0 \sim -4.2 $ days from the peak,
the epochs of our observations correspond to
$\sim 1.5 - 2.2$ days after the explosion.

The agreement between 
the luminous 2 objects in our samples (SHOOT14gp and 14or)
and PS1 faint samples is not as good as that for the faint 3 objects
(SHOOT14ha, 14jr, and 14ef).
Note that the direct comparison at the perfectly matched wavelengths is
not possible ($< 3000$ \AA\ for SHOOT14gp and 14or while
$> 4000$ \AA\ for the PS1 faint samples).
Nevertheless, SHOOT14gp and 14or show faster rises than the PS1 faint samples.
The rising rates of SHOOT14gp and 14or are
$> 3.10$ and $3.12^{+1.11}_{-0.70}$ mag day$^{-1}$, respectively
(Table \ref{tab:objects}).
On the other hand, the rising rate of the PS1 faint sample
is \dmdt $< 1.3$ mag day$^{-1}$
even at the fastest phase in the interpolated light curves
(see dashed lines in Figures \ref{fig:LCrapid} and \ref{fig:dmdt}).
The nature of these objects are discussed in Section \ref{sec:discussion}.

\section{Rising rates of transients}
\label{sec:dmdt}

Figure \ref{fig:dmdt} shows a summary of rising rate and absolute magnitudes
of our samples and other transients shown
in Figures \ref{fig:LCall}, \ref{fig:LCearly}, and \ref{fig:LCrapid}.
The figure is shown as a function of rising timescale
$\tau_{\rm rise} \equiv 1/$ \dmdt, time to have 1 mag rise.
For our objects, SN 2010aq, PS1-13arp, and the PS1 samples,
the rising rates are measured only at an interval on the rise
as there are no time-series data before the peak.
The time interval is $\Delta t \gsim 0.5$ days.
For normal SNe, for which good time-series data are available,
we measure the rising rate \dmdt\ as a function of time
(connected with lines in Figure \ref{fig:dmdt}).
In order to match the time interval with other objects,
the time interval is kept to be $\Delta t \gsim 0.5$ days.
For example, although fine time-series data are available for SN 2006aj
before the peak, we measure the rising rate from $t$=0.082 and $t$=0.541 days
from the burst ($\Delta t_{\rm rest} = 0.45$ days).
For the PS1 faint samples (PS1-10ah and PS1-10bjp),
the green dashed lines show the the rising rate
measured with $\Delta t_{\rm rest} = 0.5$ days using
the light curves interpolated with $f = (t-t_0)^2$.

In this diagram, as also discussed in Section \ref{sec:SN},
it is clear that Type Ia SN shows the fast rise
only at the very early phase with faint magnitudes.
Core-collapse SNe after a few days from the explosion
are located at the region with fainter magnitudes
and longer timescales compared with our samples.

Our samples share a region similar to SN 2010aq
and PS1-13arp, SNe with early UV detection by {\it GALEX}
\citep{gezari10,gezari15}, as expected from the comparison
in the previous sections (Figure \ref{fig:LCearly}).
The early peak of SN 2006aj also has a similar rising rate,
but it is brighter than our samples.

The PS1 luminous samples \citep{drout14}
is located at the region with brighter magnitudes and longer timescales.
On the other hand,
the PS1 faint samples are closer to the faint three objects in our samples
(SHOOT14ha, 14jr, and 14ef).
Especially, when the rising rate is measured with the interpolated
light curves to have a similar $\Delta t_{\rm rest}$ with our samples,
the brightness and the rising timescale of the PS1 faint samples
shows fairly good agreement with SHOOT14ha, 14jr, and 14ef
(see also Figure \ref{fig:LCrapid}).

\section{Discussion}
\label{sec:discussion}

The properties of our samples of rapidly rising transients are
similar to those of very early core-collapse SNe,
such as SN 2010aq, PS1-13arp, and SN 2006aj (Figure \ref{fig:dmdt}).
The faint three objects also show a similarity to
the faint population (with $> -19$ mag) of the
rapidly rising transients from PS1 \citep{drout14},
which are also interpreted as the very early phase of SNe.
For both cases, the best match is obtained when
our samples are assumed to be $\sim 1-2$ days after the explosion.

By these facts, although we do not have photometric follow-up
and spectroscopic identification of our samples,
we interpret that
the rapidly rising transients presented in this paper
are the very early phase of core-collapse SNe.
In the following sections, we discuss the nature of
the rapidly rising transients based on this interpretation.

\begin{figure*}
\begin{center}
\begin{tabular}{cc}
\includegraphics[scale=0.9]{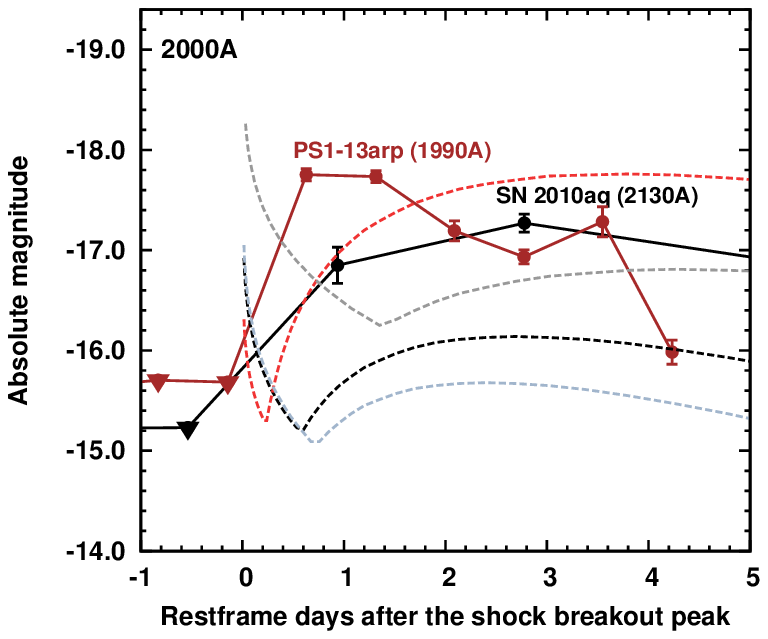} &
\includegraphics[scale=0.9]{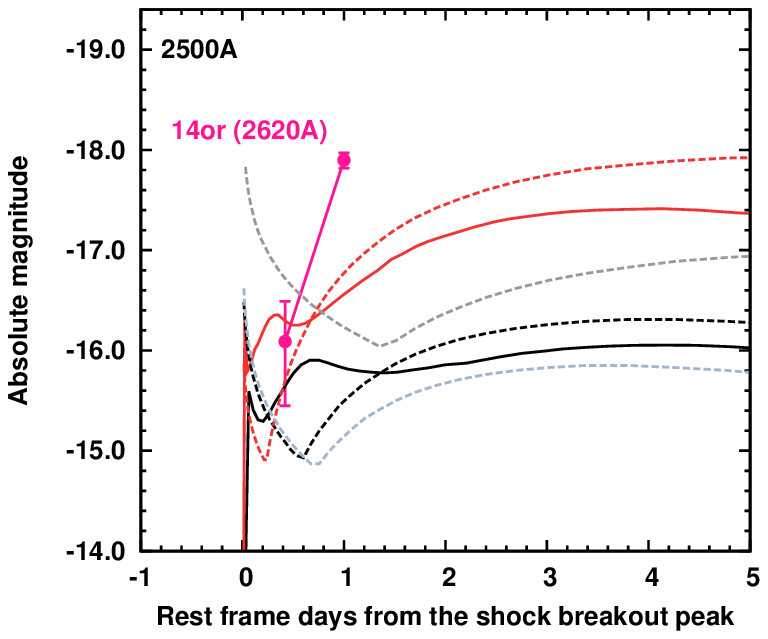} \\
\includegraphics[scale=0.9]{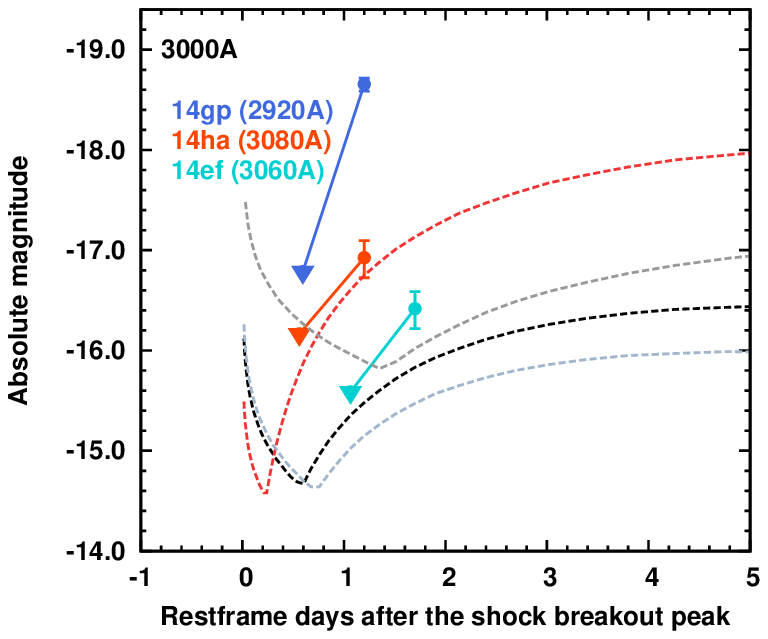} &
\includegraphics[scale=0.9]{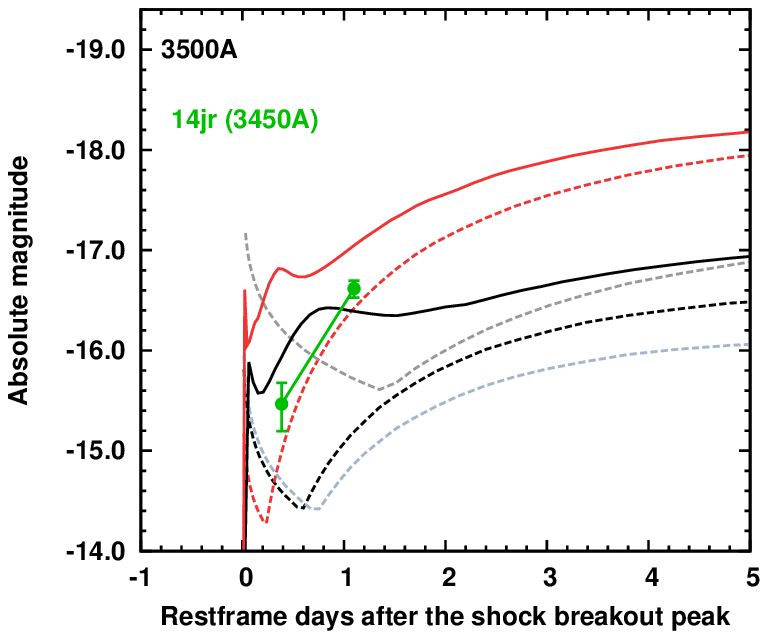} \\
\end{tabular}
\caption{
  Comparison of the light curves between
  the observed light curves and model light curves.
  The dashed lines show analytic light curve models of cooling envelope emission
  for red supergiant SNe by \citet{nakar10}:
black ($\Mej, R, E$) = ($15 \Msun, 500 \Rsun, 1.0 \times 10^{51}$ erg),
upper gray ($15 \Msun, 1000 \Rsun, 1.0 \times 10^{51}$ erg),
lower gray ($25 \Msun, 500 \Rsun, 1.0 \times 10^{51}$ erg),
and red ($15 \Msun, 500 \Rsun, 5.0 \times 10^{51}$ erg).
The black and red solid lines in the panel of 2500 \AA\ and 3500 \AA\ are
numerical models calculated with STELLA:
($\Mej, R, E$) = ($15 \Msun, 500 \Rsun, 1.2 \times 10^{51}$ erg)
and ($15 \Msun, 500 \Rsun, 4.0 \times 10^{51}$ erg), respectively.
Epochs of observed data are arbitrarily shifted to match the models.
}
\label{fig:models}
\end{center}
\end{figure*}

\begin{figure*}
\begin{center}
  \includegraphics[scale=1.5]{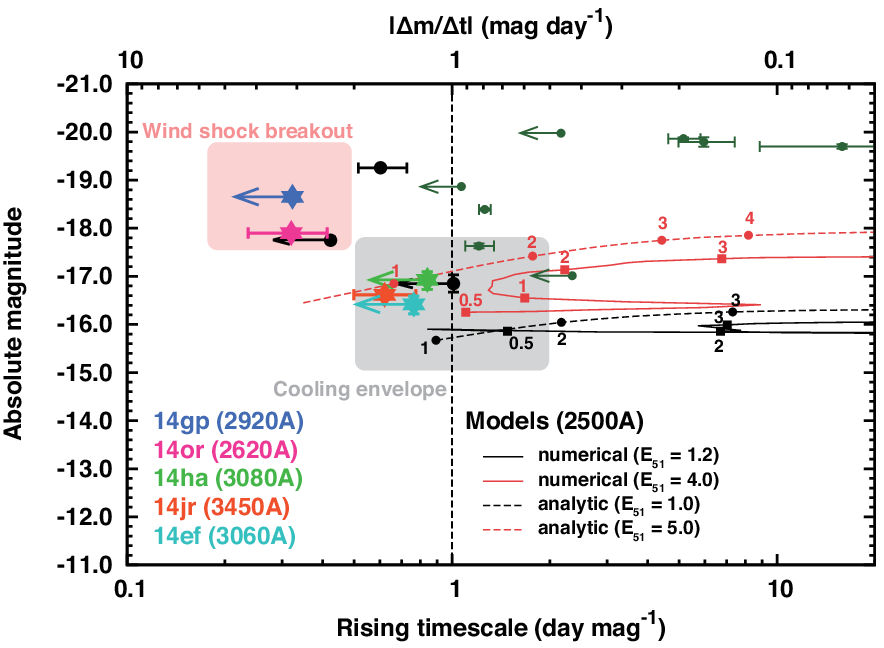}
  \includegraphics[scale=1.5]{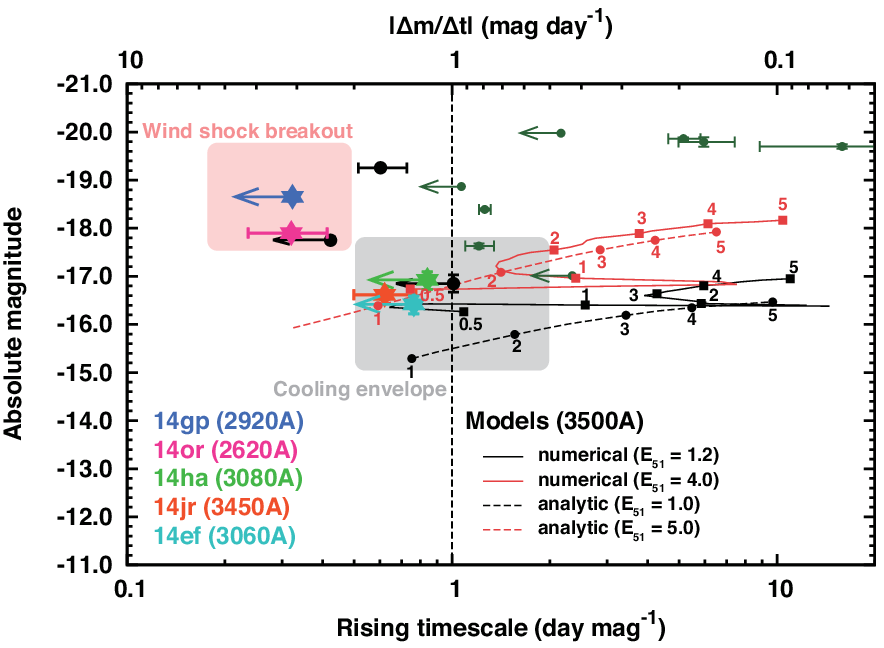}
  \caption{
    Absolute magnitude and rising timescale (as in Figure \ref{fig:dmdt})
    compared with analytic and numerical models
    of red supergiant explosion.
    Upper and lower panels show the models at 2500 \AA\ and 3500 \AA\, respectively.
    The solid and dashed lines show numerical and analytic models, respectively.
    The parameters of the models are following.
    Numerical models: black solid
    ($\Mej, R, E$) = ($15 \Msun, 500 \Rsun, 1.2 \times 10^{51}$ erg)
    and red solid ($15 \Msun, 500 \Rsun, 4.0 \times 10^{51}$ erg).
    Analytic models:
    black dashed ($\Mej, R, E$) = ($15 \Msun, 500 \Rsun, 1.0 \times 10^{51}$ erg)
    and red dashed ($15 \Msun, 500 \Rsun, 5.0 \times 10^{51}$ erg).
    The time evolution of the models are connected with lines.
    Numbers associated with dots show the epochs (in days)
    from the peak of the shock breakout.
  }
\label{fig:dmdt_model}
\end{center}
\end{figure*}

\subsection{Constraints on the event rate}

Event rates of rapidly rising transients shown in this paper
are of interest.
However, to estimate the event rates,
we need detailed information about spectral energy distribution,
light curve shape, and luminosity function,
which are not available for our samples.
Instead, we give crude constraints
on how high event rate is required
for short-timescale events to be detected with our short-period survey.

We estimate the event rates by using a method
based on $1/V_{\rm max}$ method \citep{schmidt68,eales93},
which is used for estimation of galaxy luminosity function.
The event rates of transients $R$ can be written as
$R = \sum_i R_i = \sum_i \frac{1}{p_i \tau_i V_{{\rm max},i}}$.
Here, $p_i$ is a detection efficiency ($p_i < 1$),
$\tau_i$ is the restframe time window for a rapidly rising
transient to be detected with our survey, and $V_{{\rm max},i}$ is the maximum volume
in which the transient is detectable with our survey.
The summation is taken for all the detected objects.
The difference from galaxy luminosity function is
$\tau_i$ in the denominator to take into account the fact
that transient event rate should be measured for a given time period.
As the number of samples is small,
we do not take into account redshift evolution of the event rate.

We do {\it not} correct detection efficiency
since the selection criteria are complicated:
we need spectroscopic redshift
to define the rapidly rising transients (Section \ref{sec:selection}).
Thus, we assume $p_i = 1$, so that
the analysis gives a conservative lower limit for the event rate
(see below for possible impact of this assumption).

Then, the free parameter in this analysis is only $\tau_i$.
For simplicity, we assume this parameter is the same ($\tau$)
for all the objects by neglecting different redshifts.
Here, $\tau$ means the duration for which transients show
a rapid rise with sufficient brightness
so that they are recognized as rapidly rising transients
in our survey.
For the two objects detected both on Days 1 and 2 (SHOOT14or and 14jr),
the duration of the emission is about 1.2 days in the observed frame
(0.67 and 0.86 days in the restframe, respectively),
and thus, $\tau$ is not much shorter than 1 day.
A smaller $\tau$ is not excluded for the other three objects
but they do not show clear intranight variability
for 1.6-3.1 hr in the observed frame (1.0-2.0 hr in the restframe).
Comparison with previously known transients (Section \ref{sec:LC})
and also with models (see Section \ref{sec:model}) suggest
that it is unlikely that the rising rate as high as \dmdt\ $> 1$ mag day$^{-1}$
continues for $> 2$ days in restframe with sufficient brightness.
Thus, we adopt $\tau = 1$ day as a fiducial value for all objects.

A typical 3$\sigma$ limiting magnitude for the images used for candidate
selection is $\simeq$ 26.0 mag.
We use this value for the calculation of the maximum volume $V_{\rm max}$.
In fact, for objects to be recognized as rapidly rising transients,
they should be sufficiently brighter than the limiting magnitude on Day 2.
Thus, the effective limiting magnitude for the rapidly rising transients
tends to be shallower than 26.0 mag.
Since analysis with a shallower limiting magnitude gives
a smaller maximum volume and a higher event rate,
our choice of deep limiting magnitude gives conservative estimates for the event rate.
It is noted that the extinction in the host galaxy is not corrected
and the true absolute magnitude of our samples should be brighter.
However, if the extinction for the current samples
represents an average degree of extinction, 
the estimate of $V_{\rm max}$ is not significantly affected
(\ie our estimate crudely includes the effect of extinction).

We estimate pseudo event rate for each object ($R_i$).
For example, the maximum redshift, in which our survey would have
detected SHOOT14gp, is $z_{\rm max} = 1.87$ with the limiting magnitude
of 26.0 mag using absolute magnitude of $M = -18.67$ mag and
crude K-correction (the term of $2.5\log(1+z)$) as in Section \ref{sec:LC}.
The comoving volume within this redshift
in 12 deg$^2$ survey area is $V_{{\rm max},i} = $ 0.16 Gpc$^3$.
For this object to be detected with our survey,
the required event rate should be
$R_i \simeq 1/\tau_i V_{{\rm max},i} \simeq 0.23 \times 10^{-5} (\tau/ 1 {\rm day})^{-1} $ yr$^{-1}$ Mpc$^{-3}$.
Similar analysis for SHOOT14or, 14ha, 14jr and 14ef give
$z_{\rm max} =$ 1.28, 0.70, 0.82, and 0.62, 
and the event rates are
$R_i \simeq $ 0.47, 1.9, 1.3, 2.5 $\times 10^{-5} (\tau/ 1 {\rm day})^{-1} $
yr$^{-1}$ Mpc$^{-3}$, respectively.

By summing up the pseudo rates,
the lower limit of the total event rate is
$R \simeq 6.4 \times 10^{-5} \ (\tau/ 1 {\rm day})^{-1} $ yr$^{-1}$ Mpc$^{-3}$.
It corresponds to about 9 \% of core-collapse SN rate at $z \sim 1$
(the core-collapse SN rate is $(3-7) \times 10^{-4}$ yr$^{-1}$ Mpc$^{-3}$
at $z=0-1$, \citealt{dahlen04,botticella08,li11,dahlen12}).
Note that the event rate is dominated by the less luminous object
with smaller maximum volumes.
The event rate for the two luminous events (SHOOT14gp and 14or) is
$R = 0.7 \times 10^{-5} \ (\tau/ 1 {\rm day})^{-1} $ yr$^{-1}$ Mpc$^{-3}$
($\sim 1$ \% of the core-collapse SN rate at $z \sim 1$),
while the event rate for the three faint events (SHOOT14ha, 14jr, and 14ef)
is $R = 5.7 \times 10^{-5} \ (\tau/ 1 {\rm day})^{-1} $ yr$^{-1}$ Mpc$^{-3}$
($\sim 8$ \% of the core-collapse SN rate).
It is worthy to mention that the event rate of the rapid transients
from PS1 is estimated to be $4\% - 7\%$ of core-collapse SN rate
\citet{drout14}, which is broadly consistent with our estimate.

As described above, our estimate involve crude approximation,
mainly due to (1) incompleteness of the sample,
(2) a choice of simple magnitude limit,
and (3) unknown transient duration.
To anchor a possible range of uncertainties,
we here discuss impacts of each effect.
(1) As discussed in Section \ref{sec:selection},
we could not take spectra of 6 SN candidates.
If all of them satisfy the criteria of rapid transients,
the total number of the objects is 11 instead of 5.
Actual impact to the event rate depends on their luminosity and redshifts,
but if all of them are assumed to be similar to our faint samples (with a high event rate),
the total event rate can be increased at most by a factor of about 2.2 (11/5).
(2) If a shallow magnitude limit is adopted,
it results in a smaller $V_{\rm max}$ and a higher event rate.
By adopting 25.5 mag limit, which is the possible shallowest limit
to detect SHOOT14ef,
the event rate is increased by a factor of 1.7.
(3) The effect of duration ($\tau$) is crudely expressed in a term of $\tau^{-1}$
and it can either reduce or increase the event rate.
The event rate is reduced by 2 for the duration of $\tau = 2$ days,
while it is increased by a factor of 1.4 for the duration of
$\tau = 0.7$ days (SHOOT14or).

In summary, our rate estimate is uncertain
by a factor of $\sim 2$ for reduction and $\sim 5$ for increase.
In either case, the event rate is not totally negligible compared
with the core-collapse SN rate.
Given the crude approximation in the estimate,
the true event rate can be comparable to the SN rate,
\ie the rapidly rising phase can be associated with all core-collapse SNe.

\subsection{Nature of the rapidly rising transients}
\label{sec:model}

{\it Shock breakout:}
The electromagnetic signal from SNe starts with shock breakout emission.
Shock breakout occurs when the diffusion timescale of photons
in front of the shock wave becomes
as short as the dynamical timescale \citep{falk78,klein78sbo}.
A typical duration of the shock breakout is light crossing time
of the progenitor size,
\ie $\sim 1000$ sec for a red supergiant progenitor with 500 $\Rsun$
\citep[\eg][]{matzner99,ensman92,tominaga09sbo,tominaga11}
and shorter for more compact progenitors.

Timescales of shock breakout emission are much shorter
than the observed timescale for SHOOT14or and 14jr,
which are detected both on Days 1 and 2 (0.55-0.72 days in restframe).
Therefore, they can not be shock breakout emission.
On the other hand, the other three objects
(SHOOT14gp, 14ha, and 14ef) are not detected on Day 1,
and thus, the possibilities of the shock breakout are not ruled out.
However, they do not show significant
intranight variability within 1.0, 1.1, and 2.0 hr (restframe)
on Day 2, respectively, and there is no supportive signature for shock
breakout interpretation (see \citealt{tominaga15} for
the detection of a transient with an extremely rapid decline,
which is interpreted to be shock breakout emission).

{\it Cooling envelope emission:}
Following shock breakout emission,
SNe show emission from cooling envelope
\citep{waxman07,chevalier08,nakar10,rabinak11}.
This phase is believed to have been detected for SNe with very early detection,
such as SNe 2006aj and 2008D
\citep[but see \citealt{bersten13} for caveats on SN 2008D]{waxman07,soderberg08,modjaz09,chevalier08,nakar15}.
The early UV detection of SN 2010aq (Figure \ref{fig:LCearly})
is also interpreted as a cooling emission \citep{gezari10}.
\citet{drout14} also showed that,
among their rapid transients from PS1,
the faint objects such as PS1-10ah 
can be interpreted as the cooling envelope emission.
In addition to these very early detection,
the tail of the cooling phase is sometimes observed 
in some other SNe, such as SNe 1993J, 1999ex, and 2011dh,
at later phases \citep[\eg][]{lewis94,richmond94,stritzinger02,arcavi11,marion14}.

Figure \ref{fig:models} shows light curves of cooling envelope emission
for red supergiant cases by \citet{nakar10},
compared with light curves of our samples, SN 2010aq, and PS1-13arp.
We divide these objects into 4 classes according to
effective restframe wavelengths
(2000, 2500, 3000, and 3500 \AA).
The black dashed lines show the fiducial model with
the ejecta mass $\Mej = 15 \Msun$, progenitor radius $R = 500 \Rsun$,
and explosion energy $E = 1.0 \times 10^{51}$ erg.
Other lines show models with different mass, radius, and energy:
upper gray dashed line ($\Mej, R, E$)
= ($15 \Msun, 1000 \Rsun, 1.0 \times 10^{51}$ erg),
lower gray dashed ($25 \Msun, 500 \Rsun, 1.0 \times 10^{51}$ erg),
and red dashed ($15 \Msun, 500 \Rsun, 5.0 \times 10^{51}$ erg).
The epochs of observed data are arbitrarily shifted to match the models.
The brightness of observed samples is consistent or brighter than
the red supergiant models.
Since the cooling envelope emission from explosions of more compact
progenitor tend to be fainter than red supergiant case
in UV at $\sim 1$ day \citep{nakar10},
models with blue supergiant or Wolf-Rayet star progenitors
do not give better agreement.

The light curve of SHOOT14jr
is qualitatively consistent with a model of cooling envelope emission.
SHOOT14ha and 14ef can also be explained by the models,
although they are detected only Day 2.
Since the cooling envelope emission peaks at a epoch
when $h \nu \sim 3kT$ is fulfilled,
the spectral peak at the rising phase is located at shorter wavelengths
than the observed wavelengths.
This is also consistent with the blue color of our objects.
Note that comparison with the models suggest an explosion energy
higher than $1.0 \times 10^{51}$ erg.
In addition, due to possible extinction in the host galaxies,
the true absolute magnitudes of our objects can be even brighter.
These situations are also the case for SN 2010aq, where a model brighter than
our fiducial model by $1.5$ mag gives the best match with the observed data
without host extinction correction \citep{gezari10}.

To understand possible varieties in the models,
we also show selected numerical models for the early phase of Type IIP SNe.
The models are calculated with the
multigroup radiation hydrodynamics code STELLA \citep{blinnikov06}.
For the purpose of parametric studies,
quasi-polytrope pre-SN models are constructed
in hydrostatic equilibrium by assuming the solar metallicity
and a power-law dependence of the temperature on the density
as in \citet{baklanov05,baklanov15}.    
In Figure \ref{fig:models}, 
magnitudes in {\it Swift} $uvw1$ and $u$-filters are shown
in the panels of 2500 \AA\ and 3500 \AA\ data.
Black and red solid lines show 
the models with similar parameters to those for analytic models:
($\Mej, R, E$) = ($15 \Msun, 500 \Rsun, 1.2 \times 10^{51}$ erg)
and ($15 \Msun, 500 \Rsun, 4.0 \times 10^{51}$ erg), respectively.
Although there are some discrepancy between analytic and numerical models,
the trend is similar:
SHOOT14jr can be consistent with models
while SHOOT14or is brighter and faster than the models.

Figure \ref{fig:dmdt_model} shows the rising timescales and
absolute magnitudes (as in Figure \ref{fig:dmdt})
compared with those of analytic (dashed) and numerical (solid) models.
The black and red lines show the fiducial models
and models with a higher energy.
As also shown in Figure \ref{fig:models},
the light curve models are consistent with the faint three
objects in our samples at $\lsim 1-2$ days after the shock breakout.

In summary, the three faint objects (SHOOT14ha, 14jr, and 14ef)
out of our five samples
are interpreted to be the cooling envelope emission of
red supergiant explosion.
The epochs of our detection is likely to be $\lsim 1-2$ days
after the shock breakout.

{\it Shock breakout from dense wind:}
SHOOT14gp and SHOOT14or, two luminous objects in our samples,
are brighter and faster than the cooling envelope models.
In fact, this difficulty is also found
for the case of PS1-13arp, and
\citet{gezari15} suggested that it is shock breakout
from a dense wind since the luminosity of the shock breakout
from the wind can be more luminous
than cooling envelope emission by factor of $\gsim 10$
\citep{ofek10,moriya11,chevalier11,balberg11}.
  
For the shock breakout from the wind,
the timescale to the peak luminosity reflects the
diffusion timescale in the wind,
$t_p = 6.6 \ (\kappa/0.34\ {\rm cm^2\ g^{-1}}) (\dot{M}/10^{-2}\ {\rm \Msun\ yr^{-1}}) (v_{\rm wind} / 10\ {\rm km\ s^{-1}})$ days \citep{chevalier11},
where $\dot{M}$ and $v_{\rm wind}$ is the mass loss rate and wind velocity, respectively.
For our samples, the time to the peak is not tightly constrained,
but it is longer than $0.55$ days for SHOOT14or.
Therefore, the required mass loss rate is the order of
$10^{-3}\ \Msun $ yr$^{-1}$ for the wind velocity of $v_{\rm wind} = 10$ \kms.
A typical epoch when such a mass loss rate is required is
$t_{\rm wind} \sim 2.7 \ (v_{\rm SN}/10,000 \ {\rm km\ s^{-1}}) (v_{\rm wind}/10 \ {\rm km\ s^{-1}})^{-1} (t_{\rm SN}/ 1\ {\rm day})$ years before the explosion,
where $v_{\rm SN}$ and $t_{\rm SN}$ are shock velocity of SN
and observed time after the explosion, respectively.

The inferred mass loss rate is
as high as enhanced, episodic mass loss rate
estimated for VY Canis Majoris
\citep[$(1-2) \times 10^{-3}\ \Msun$ yr$^{-1}$,][]{smith09RSG},
and higher than that typically estimated for red supergiants,
$\dot{M} \lsim 10^{-4} \Msun$ \citep{vanloon05,mauron11}.
If our interpretation is the case, our study implies that
$\gsim 1 \%$ of massive stars can have such a high mass loss rate
at the very end of the stellar evolution
(\ie a few years before the explosion).

\citet{drout14} also suggested that the PS1 luminous samples
are the shock breakout from the wind.
PS1 luminous samples show longer timescale
than those for our two luminous samples and PS1-13arp
(Figure \ref{fig:dmdt}).
This may be understood as the different mass loss rates of the wind:
the PS1 luminous samples require a higher mass loss rates
$\sim 10^{-2}\ \Msun $ yr$^{-1}$ \citep{drout14}.

\section{Conclusions}
\label{sec:conclusions}

We perform a high-cadence transient survey using Subaru/HSC.
In the observations of two continuous nights,
we detected five rapidly rising transients
at $z =0.384 - 0.821$ with the rising rate faster than
1 mag per 1 day in restframe (\dmdt > 1 mag day$^{-1}$).
The absolute magnitudes of the five objects
range from $-16$ to $-19$ mag in the restframe near-UV wavelengths,
and they all show blue colors, $g-r \lsim -0.2$ mag.

To our knowledge,
the rising rate and brightness of our samples are
the most similar to those of the very early phase
($<$ a few days after the explosion) of core-collapse SNe,
such as SN 2010aq and PS1-13arp detected by
{\it GALEX} at the very early phases \citep{gezari10,gezari15},
and the faint population of rapid transients from PS1 \citep{drout14}.
A conservative estimates suggest that the event rate of
rapidly rising transients is $\gsim$ 9 \% of core-collapse SN rates,
assuming a typical duration of the fast rising phase
in the near-UV wavelengths to be 1 day.
The true event rate can be comparable to the core-collapse SN rate.

Although spectroscopic identification is not available,
the rapidly rising transients presented in this paper
are interpreted to be the
very early phase of core-collapse SNe.
The observed light curves of faint three objects
(SHOOT14ha, 14jr, and 14ef)
are qualitatively consistent with
the cooling envelope emission from the explosion of red supergiants.
The comparison with the analytic and numerical models shows
that the epochs of our
observations correspond to $\lsim 1-2$ days after the shock breakout.

The other two luminous objects (SHOOT14gp and 14or) are brighter
and faster than
the expectation of the cooling envelope models.
We interpret that they are 
shock breakout emission from the dense wind,
as also suggested for PS1-13arp.
The required mass loss rate is $\sim 10^{-3}\ \Msun$ yr$^{-1}$.
The event rate of these luminous events is higher than $\sim 1 \%$ of
core-collapse SN rate.
Therefore, if our interpretation is correct,
it implies that more than $\sim 1 \%$ of massive stars
can experience such a strong mass loss at a few years before the explosion.

\acknowledgments

We thank the anonymous referee for
constructive comments that improved the paper.
This research was in part supported by 
Grants-in-Aid for Scientific Research of JSPS
(23224004, 23740157, 24740117, 25800103, 26400222, 15H02075, 15H05440),
MEXT (25103515, 15H00788),
the World Premier International Research Center Initiative, MEXT, Japan,
the research grant program of Toyota foundation (D11-R-0830),
and the RFBR-JSPS bilateral program.
S.B. and P.B. are supported in the work on STELLA code by the Russian
Science Foundation Grant No. 14-12-00203.
TJM is supported by JSPS Postdoctoral Fellowships for Research Abroad
(26\textperiodcentered 51).
Support for HK is provided by the Ministry of Economy, Development,
and Tourism's Millennium Science Initiative through grant IC120009,
awarded to The Millennium Institute of Astrophysics, MAS.
HK acknowledges support by CONICYT through FONDECYT grant 3140563.
This paper makes use of software developed for the LSST.
We thank the LSST Project for making their code available
as free software at http://dm.lsstcorp.org.


\end{document}